\documentclass[aps,preprint,preprintnumbers,amssymb,amsfonts,groupedaddress]{revtex4} %subeqn,

\usepackage{graphicx}
\usepackage{dcolumn}% Align table columns on decimal point
\usepackage{bm}% bold math
\usepackage{amsmath}
\usepackage{color}

\newcommand{\bmath}{\begin{mathletters}}
\newcommand{\emath}{\end{mathletters}}
\newcommand{\be}{\begin{eqnarray}}
\newcommand{\ee}{\end{eqnarray}}
\newcommand{\ba}{\begin{array}}
\newcommand{\ea}{\end{array}}

\newcommand{\no}{\nonumber}

\newcommand{\bt}{\beta}
\newcommand{\de}{\delta}

\newcommand{\Ga}{\Gamma}

\newcommand{\e} {\varepsilon}
\newcommand{\vare}{\varepsilon}

\newcommand{\h}{\hbar}
\newcommand{\pr}{\prime}

\newcommand{\calK}{\mathcal K}
\newcommand{\calL}{\mathcal L}
\newcommand{\calU}{\mathcal U}
\newcommand{\calW}{\mathcal W}

\newcommand{\eq} {\mathrm{eq}}
\newcommand{\Tr} {\mathrm{Tr}}
\newcommand{\Real}{\mathrm{Re}}
\newcommand{\Imag}{\mathrm{Im}}
\newcommand{\rmP} {\mathbf{P}}

\newcommand{\rmC} {\mathbf{C}}

\begin{document}
\title{Higher-Order Kinetic Expansion of Quantum Dissipative Dynamics: Mapping Quantum Networks to Kinetic Networks}
\author{Jianlan Wu$^{1,2}$}
\author{Jianshu Cao$^2$\footnote{E-mail: jianshu@mit.edu.}}
\affiliation{$^1$Physics Department, Zhejiang University, 38 ZheDa Road, Hangzhou, Zhejiang, 310027, China, \\
$^2$Department of Chemistry, MIT, 77 Massachusetts Ave, Cambridge, MA, 02139, USA}
\date{today}

\begin{abstract}
We apply a new formalism to derive the higher-order quantum kinetic expansion (QKE)
for studying dissipative dynamics in a general quantum network coupled with an arbitrary thermal bath.
The dynamics of system population is described by a time-convoluted kinetic equation,
where the time-nonlocal rate kernel is systematically expanded on the order of off-diagonal
elements of the system Hamiltonian. In the second order, the rate kernel recovers the
expression of the noninteracting-blip approximation (NIBA) method.
The higher-order corrections in the rate kernel
account for the effects of the multi-site quantum coherence and the bath relaxation.
In a quantum harmonic bath, the rate kernels of different orders are analytically derived.
As demonstrated by four examples, the higher-order QKE can reliably
predict quantum dissipative dynamics, comparing well with the hierarchic equation approach.
More importantly,
the higher-order rate kernels can distinguish and quantify distinct nontrivial quantum
coherent effects, such as long-range energy transfer from quantum tunneling and quantum interference
arising from the phase accumulation of interactions.

\end{abstract}
%\pacs{xxx}

\maketitle

\section{Introduction}

Quantum dissipation plays a key role in understanding quantum dynamic processes.
The interaction between a quantum system and its surrounding environment causes
an irreversible loss of the energy and coherence of the quantum system.
The relaxation and decoherence times are the limiting factor of the quantum
computation and quantum information~\cite{Breuer2002}. %Divincenzo2000:Nature
In the Caldeira-Leggett model, the change of the dissipation strength
can interpret quantum tunneling and localization in macroscopic systems~\cite{Leggett1981:PhysRevLett,Leggett1987:RMP}.
For many years, the solvent modulation in chemical reactions and quantum transport processes
have attracted a lot of attentions~\cite{May2004,Nitzan2006}.
Incorporated with the description of the solvent reorganization,
the Marcus theory is able to explain essential features of
electron transfer~\cite{Marcus1964:ARPC}.
In the recent two-dimensional (2D) electronic light spectroscopy,
long-lived quantum coherence and wavelike dynamics have been
found in natural light-harvesting protein complexes~\cite{Engel2007:Nature} %,Harel2012:PNAS,Collini2010:Nature}
and organic conjugated polymers~\cite{Collini2009:Science},
which also triggers studies on the energy transfer optimization from
the dissipation induced by the protein environment~\cite{JLWu2010:NJP,Moix2011,JLWu2012:JCP,JLWu2012:Orthogonal,Plenio2008:NJP,Rebentrost2009:NJP,Brumer2011:JPCL}.
To understand the nontrivial effect of quantum coherence in the energy transfer,
we need to study the underlying quantum dissipative dynamics
beyond the conventional F\"{o}rster resonance energy transfer (FRET) theory~\cite{Forster1948}.

An accurate and reliable approach to compute quantum dissipative dynamics
is a long-lasting but difficult theoretical problem.
A large number of methods have been developed
under many different frameworks, such as the Nakajima-Zwanzig projection
operator~\cite{Nakajima1958:PTP,Zwanzig1960:JCP}, the Feynman-Vernon influence functional approach~\cite{Feynman1963:AnnPhys},
and quantum stochastic noises formulation~\cite{Mak1998:PhysRevLett,Shao2004:JCP,Cao1996:JCP}. %Stockburger2002:PhysRevLett%Strunz1999:PhysRevLett
The second-order perturbation methods, such as Fermi's golden rule rate,
Redfield equation~\cite{Redfield1957:IBMJ}, generalized Bloch-Redfield equation~\cite{Cao1997:JCP,JLWu2010:NJP},
and noninteracting-blip approximation (NIBA)~\cite{Leggett1987:RMP},
are derived in the limit of either a weak or strong system-bath interaction.
%Despite their great success, these methods
%are limited in their special parameter regimes
 %a qualitative or quantitative  prediction
%and can fail elsewhere.
In the variational polaron approach~\cite{Harris1983:JCP},
a self-consistent reference can partially
improve the prediction of the second-order perturbation.
With a classical bath, the Haken-Strobl-Reineker (HSR) model~\cite{Haken1972:ZPhysik,Haken1973:ZPhysik,Silbey1976:ARPC,cao2009:JPCA}
and other quantum-classical mixed methods~\cite{Berkelbach2012:JCP,Aghtar2012:JCP,Tully1971:JCP,Kapral2012:JCP} %Micha1983:JCP,
describe dissipative dynamics at high temperatures.
The dissipative dynamics under a quantum harmonic bath can be
evaluated by many sophisticated methods, such as
the semiclassical initial value representation (SC-IVR)~\cite{Miller1998:FaradayDiscuss},  %Miller2001:JPhysChemA,Miller2010:JPCL
the iterative linearized density matrix (ILDM) propagation~\cite{Huo2010:JCP},
the quasi-adiabatic propagator path integral (QUAPI)~\cite{Makri1995:JChemPhys1}, %,Makri1995:JChemPhys2
the path integral Monte Carlo~\cite{Mak1996:AdvChemPhys,Muhlbacher2004:JChemPhys}, etc. %Suzuki1993,Muhlbacher2004:JChemPhys,Muhlbache2012:JPCB}, etc.
Despite their successes, these %sophisticated deterministic or stochastic
methods can be numerically expensive and becomes difficult
for long-time dynamics. %due to accumulated errors.
If the time correlation of the harmonic bath can be expanded
as a sum of exponentially decaying functions, the hierarchy equation approach can
%in principle
accurately predict quantum dissipative dynamics by expanding over auxiliary
fields~\cite{Tanimura1989:JPSJ,Tanimura2005:JPSJ,Ishizaki2009:JCP2,Shao2004:CPL,Yan2005:JCP}. %Ishizaki2009:JCP1, Ishizaki2009:PNAS,
However, the hierarchy equation is numerically difficult for large-scale systems and the bath with a long-tail correlation,
and converges slowly for strong system-bath interactions and low temperatures.

Hopping kinetics of the Fermi's golden rule rate is often
considered as a `classical' description of quantum dissipative
dynamics, although the two-`site' quantum coherence
is included in the rate expression~\cite{cao2009:JPCA,JLWu2012:JCP}.
The interesting quantum phenomena beyond the
second-order hopping kinetics can
be attributed to nontrivial quantum effects of multi-`site' coherence, e.g.,
long-range transfer (tunneling) and quantum interference.
The temporal correlation of bath is also crucial in
understanding the full quantum dynamics.
The higher-order bath relaxation effect is caused by the deviation
from the system-bath factorized reference state.
In addition, the second-order hopping kinetics
cannot predict the detailed balance of quantum dynamics, i.e.,
the Boltzmann equilibrium distribution
including both the system and the bath~\cite{Moix2012:PRB,Moix2012:JCP}.
The comparison of the second-order hopping kinetics and the full quantum dynamics
in our previous paper~\cite{JLWu2012:JCP} has
revealed the integrated behavior of the above effects, together with the flux network
analysis. To distinguish and quantify nontrivial quantum effects,
%and bath relaxation effect,
we need a systematic expansion procedure to
calculate every higher-order correction beyond the second-order hopping kinetics,
which is almost impossible in many sophisticated theoretical methods discussed above.
%Although the expansion technique is inherited in the hierarchy equation,
%various higher-order corrections are mixed in the same expansion order.
%Reference~\cite{cao2009:JPCA} used a stationary approximation for the coherent term
Following the stationary approximation for the coherent term,
the kinetic mapping of quantum dynamics  allows us to identify the
multi-site quantum coherence term by term in the HSR model~\cite{cao2009:JPCA}.
However, the theoretical method for a general quantum bath is still missing.

To address the above concerns, we will apply a general non-Markovian
quantum kinetic equation, where the time-nonlocal rate kernel is obtained
by a systematic expansion approach. This higher-order quantum kinetic expansion
(QKE) method presents a rigorous mapping from a quantum network to a kinetic network,
which helps us to identify nontrivial quantum effects and bath relaxation beyond the
traditional classical description. In addition, the higher-order QKE is expected to serve as a
numerically reliable method to calculate the  population dynamics.
For simplicity, we focus on a product state between the system and the bath at zero time,
and assume that quantum system is initially prepared in the population subspace
without quantum coherence.
This initial incoherent assumption is acceptable for
an initially localized (quasi-) particle in the quantum transport process,
e.g., an exicton after absorbing incoherent sunlight in a natural light-harvesting
protein complex~\cite{May2004}.
Accordingly, the bath induced fluctuation is assumed on the diagonal elements
of the system Hamiltonian. Since the system is not defined in its eigenbasis
(the system Hamiltonian is not diagonal), the diagonal fluctuation can still lead to
both relaxation and decoherence.
As we discussed, many theoretical approaches require the presumption of
the harmonic bath, which is often considered
as a good approximation under many circumstances.
%It is valid when bath modes oscillate mainly
%around their equilibrium position, e.g., ultrafast quantum dynamics where
%a large-scale change of bath is rare in the relevant time scale.
In complex environments such as a protein backbone,
anharmonicity however could be relevant even in a  fast energy transfer process. %even in ultrafast quantum dynamics.
Here, the higher-order QKE is free of the harmonic bath presumption,
although the additional numerical implementation is required in the anharmonic bath.

An essential part of our theory is the systematic
expansion of time-nonlocal kinetic rate kernels. For the two-site system,
the bath relaxation effect was calculated in electron transfer
following a term-by-term comparison in integrated population between
microscopic expansion of quantum dynamics and a formally exact %assigned
non-Markovian kinetic equation~\cite{Cao2000:JCP2}.
Here we will extend this procedures to a multi-site system.
Although some of the higher-order corrections have been studied
in the past~\cite{Mukamel1989:JCP,Skinners1991:JCP,Reichman1996:JCP,Weiss1994:PRE}, %Reichman1997:PhysRevE, Mukamel1988:JPC,
the higher-order QKE in this paper provides a systematic %new
formalism of obtaining the general expression of the time-nonlocal
rate kernels and unify the bath relaxation and the
multi-site coherence under the same framework for a multi-site quantum network.

The paper is organized as follows: In Sec.~\ref{sec2}, we will integrate the time evolution of
quantum coherence and project the Liouville equation to be a closed dynamic equation of
system population. In Secs.~\ref{sec3}, we will develop the non-Markovian
higher-order quantum kinetic expansion method. The time-nonlocal kinetic rate kernel is generalized
from the simplest second-order, i.e., the NIBA expression, to an arbitrary $k$-th order.
The time correlation function formalism in the Hilbert space provides
rigorous expressions of rate kernels in an arbitrary bath.
In Sec.~\ref{sec4}, we will focus on the harmonic bath and
apply the displacement operator and the cumulant expansion to derive the analytical expressions of rate kernels.
In Sec.~\ref{sec5}, the higher-order QKE is applied to four model systems for its reliability. In addition to its numerical accuracy,
we identify the bath-induced slow-down in the quantum transport rate,
and two nontrivial quantum coherent effects, quantum tunneling and quantum phase interference.
In Sec.~\ref{sec6}, we will conclude and discuss the higher-order QKE method.

\section{Population Dynamics Projected from the Liouville Equation}
\label{sec2}

For an arbitrary open quantum network, the total Hamiltonian is %formally
written as $H = H_S+H_B+H_{SB}$,
where $H_S$ and $H_B$ denote the bare system Hamiltonian and the bath Hamiltonian, respectively.
The interaction between the system and the bath is described by $H_{SB}$.
The bare system Hamiltonian $H_S$ is defined in its $N$-dimensional ($N$-D) Hilbert space.
In the single-excitation manifold of a Frenkel exiciton system,
the $n$-th basis of the Hilbert space, $|n\rangle = |0, \cdots, 1, 0, \cdots\rangle$,
represents a combination of one excitation state at the $n$-th local chromophore site
and the ground state at all the other chromophore sites~\cite{May2004}.
The total Hamiltonian $H$ can be also expanded in the $N$-D system Hilbert space using
$|n\rangle|b\rangle= |0, b_1; \cdots; 1, b_n; 0, b_{n+1}; \cdots\rangle$,
where $|b_n\rangle=|b_{n1}, b_{n2}, \cdots, b_{nM_n}\rangle$ is the complete
basis set of  $M_n(\rightarrow\infty)$ bath modes sorrounding $|n\rangle$.
Here we consider a bath-induced fluctuation, $H_{SB; n}=\sum_{b,b^\pr}H_{SB; n, b; n, b^\pr}|b\rangle\langle b^\pr|$,
over each diagonal element ($\e_n$) of $H_{S}$.
The fluctuation over off-diagonal elements ($J_{mn}$) of $H_S$ is not included in our current model;
this approximation is often applied in the study of energy and charge transfer~\cite{May2004,Nitzan2006}.
The total Hamiltonian is given as
\be
H &=& \sum_{n} H_n |n\rangle \langle n| +\sum_{m\neq n} J_{mn} |m\rangle \langle n|,
\label{eq001}
\ee
where $H_n = \varepsilon_n + H_{B}+H_{SB; n}$ is implicitly a quantum operator of bath.

The time evolution of the total density matrix $\rho(t)$
for the system-bath Hamiltonian is governed by the Liouville equation,
$\dot{\rho} =-i\mathcal L \rho$,
where $\mathcal L=[H, \cdots]$ is the Liouville superoperator.
The Planck constant $\h$ is treated as a unit throughout this paper.
With respect to
the system basis $\{|n\rangle\}$, we divide $\rho$
into two sets: population $\rho_{\rmP}=\{\rho_{nn}\}$ and coherence $\rho_{\rmC}=\{\rho_{mn(\neq m)}\}$.
Each element, $\rho_{nn}$ or $\rho_{mn}$, is a quantum operator of bath and implicitly includes
information of the entangled system and bath, i.e.,
$\rho_{nn}=\sum_{b,b^\pr}\rho_{n,b; n, b^\pr}|b\rangle\langle b^\pr|$ and
$\rho_{mn}=\sum_{b,b^\pr}\rho_{m,b; n, b^\pr}|b\rangle\langle b^\pr|$.
In this paper, we will derive our theory using both Hilbert and Liouville frameworks.
To distinguish notations in these two frameworks,
we will use `state' to specify a density state
(population and coherence) in the Liouville space, unless otherwise explained.
The wavefunction basis of the Hilbert space will be referred as `site',
consistent with the single-excitation manifold in the multi-site exciton network.
The orginal Liouville  equation is divided into two coupled equations,
\begin{subequations}
\begin{align}
%\be
\dot{\rho}_{\rmP} = -i \mathcal L_{\rmP} \rho_{\rmP} -i \mathcal L_{\rmP \rmC} \rho_{\rmC}, &\label{eq002a} \\
\dot{\rho}_{\rmC} = -i \mathcal L_{\rmC} \rho_{\rmC}-i\mathcal L_{\rmC \rmP} \rho_{\rmP}, &\label{eq002b}
%\ee
\end{align}
\end{subequations}
where the two subscripts $\rmP$ and $\rmC$ denote population and coherence of the system, respectively.
The total Liouville superoperator is expressed in a block matrix form,
\be
\calL = \left(\ba{cc} \calL_{\rmP} & \calL_{\rmP\rmC} \\ \calL_{\rmC\rmP} & \calL_{\rmC} \ea\right).
\label{eq003}
\ee
Next the coherence vector is integrated to yield
\be
\rho_{\rmC}(t) = \calU_{\rmC}(t)\rho_{\rmC}(0) -i \int_0^t d\tau_2\int_0^t d\tau_1 \de_{\tau_1+\tau_2, t} \calU_{\rmC}(\tau_2)  \calL_{\rmC\rmP} \rho_{\rmP}(\tau_1),
\label{eq004}
\ee
where $\mathcal U_\rmC(t) = \exp(-i \calL_\rmC t)$ is the time evolution matrix of coherence in the Liouville space.
For simplicity, we assume zero initial coherence, $\rho_{\rmC}(0)=0$,
so that the first term on the right hand side
of Eq.~(\ref{eq004}) vanishes.
More complicated initial conditions will be left in the future.
In the two-`site' system, $\calL_{\rmC}$ and $\calU_{\rmC}$
are diagonal in the system basis, i.e., $\calL_{\rmC; 12, 21} = 0$ and $\calU_{\rmC; 12, 21}=0$.
In the $N(>2)$-`site'  systems,  $\calL_{\rmC}$ is no longer %block
diagonal, but diagonal and off-diagonal elements
might be distinguished by their orders of magnitude,
e.g., $|\calL_{\rmC; mn, mn}|\gg |\calL_{\rmC; mn, m^\pr n^\pr(\neq m n)}|$,  in the strong damping limit.
We express $\calL_\rmC$ as a sum of the diagonal matrix
$\calL^{(0)}_{\rmC; mn, m^\pr n^\pr}=\calL_{\rmC; mn, mn}\de_{m^\pr, m}\de_{n^\pr, n}$,
and the remaining term, $\calL^{(1)}_{\rmC} =\calL_{\rmC}-\calL^{(0)}_{\rmC}$.
As shown in Sec.~\ref{sec3d}, the separation of $\calL^{(0)}_{\rmC}$
and $\calL^{(1)}_{\rmC}$ is equivalent to the separation of $H_n$ and $J_{mn}$.

Expanding the coherence vector $\rho_\rmC$ in the order of $\calL^{(1)}_\rmC$
and substituting the result into Eq.~(\ref{eq002a}), we obtain a closed time evolution equation of
population,
\be
\dot{\rho}_{\rmP}(t) %=  -i \mathcal L_{\rmP}\rho_{\rmP}(t)  - i \mathcal L_{PC}\rho_C(t) \no \\
 =  -i \mathcal L_{\rmP}\rho_{\rmP}(t) + \sum_{k=2}^\infty
\int  (-i)^{k}\mathcal L_{\rmP\rmC} \calU^{(0)}_{\rmC}(\tau_{k}) \mathcal L^{(1)}_{\rmC}  \calU^{(0)}_{\rmC}(\tau_{k-1})\cdots
\mathcal L^{(1)}_{\rmC} \calU^{(0)}_{\rmC}(\tau_2)\mathcal L_{\rmC \rmP} \rho_{\rmP}(\tau_1).
\label{eq005}
\ee
Equation~(\ref{eq005}) shows that a population flow always passes
intermediate quantum coherence states in the Liouville space of density states.
In the two-site system, population and coherence  states appear subsequently,
since the direct interconversion is not allowed between
the two coherence states, $\rho_{12}$ and $\rho_{21}$ ($\calL_{\rmC; 12, 21} = 0$)~\cite{Leggett1987:RMP}. %,Weiss1999:book}.
In the $N(>2)$-site system, the direct interconversion between different coherence states is allowed
by a nonzero `interaction', $\calL^{(1)}_{\rmC}$, from the multi-site quantum coherence.
The interactions responsible for the transition between coherence and population
states are $\calL_{\rmP\rmC}$ and $\calL_{\rmC\rmP}$.
All the three terms, $(\calL^{(1)}_\rmC,~\calL_{\rmP\rmC},~\calL_{\rmC\rmP})$,
arise from the off-diagonal
elements $J_{mn}$ of the bare system Hamiltonian, and will be counted together in the expansion order
of the final quantum kinetic equation (QKE).
For conciseness, we introduce the $k$-th order time-nonlocal population transition matrix,
\be
\calW^{(k)}&=& -(-i)^{k}\mathcal L_{\rmP\rmC} \calU^{(0)}_{\rmC}(\tau_{k}) \mathcal L^{(1)}_\rmC  \calU^{(0)}_{\rmC}(\tau_{k-1})\cdots
\mathcal L^{(1)}_{\rmC} \calU^{(0)}_{\rmC}(\tau_2)\mathcal L_{\rmC\rmP},
\label{eq006}
\ee
where $k$ is the total number of $\calL$ terms, including $(\calL^{(1)}_\rmC,~\calL_{\rmP\rmC},~\calL_{\rmC\rmP})$.
Using the complete  transition matrix, $\calW = \calW^{(2)}+\calW^{(3)}+\calW^{(4)}+\cdots$,
we formally rewrite the time evolution equation of population as
\be
\dot{\rho}_\rmP(t) &=& -i \mathcal L_{\rmP}\rho_{\rmP}(t) - \calW\ast \rho_\rmP,
\label{eq007}
\ee
where the symbol $\ast$ represents a general time convolution form,
\be
X\ast Y = \int_0^t d\tau_1  \cdots d\tau_i d\tau_{i+1} \cdots d\tau_j X(\tau_1, \cdots, \tau_i)
Y(\tau_{i+1}, \cdots, \tau_j) \de_{\tau_1+\cdots+\tau_i+\tau_{i+1}+\cdots+\tau_j, t},
\ee
for two arbitrary functions $X(\tau_1, \cdots, \tau_i)$ and
$Y(\tau_{i+1}, \cdots, \tau_j)$ of time.
Equation~(\ref{eq007}) is equivalent to
the projection of the original Liouville equation
onto the population subspace without averaging over bath.
A pratical computation of the reduced system population dynamics
relies on further simplifications  introduced in Sec.~\ref{sec3}.

\section{Non-Markovian Higher-Order  Quantum Kinetic Equation}
\label{sec3}
\subsection{Local Born Approximation and Multi-Site Quantum Coherence}
\label{sec3a}

In microscopic quantum systems, a useful physical observation is the time scale
separation between different degrees of freedom.
In this subsection, we assume that each local bath $b_n$ can instantaneously relax to
its Boltzmann equilibrium density state,
$\rho^{\eq}_{b_n} = \exp(-\bt H_n)/\Tr_b\{\exp(-\bt H_n)\}$,  at any moment.
Thus, the $n$-th element of the transient total population vector is written in
a product form, $\rho_{\rmP; n}(t) = P_n(t) \rho^{\eq}_{b_n}$~\cite{Cao2000:JCP2}.
This locally fast bath (Born) approximation
is different from $\rho(t)=\rho_S(t)\rho^\eq_b$ and $\rho^\eq_b \propto\exp(-\bt H_B)$ in the Redfield
equation~\cite{Redfield1957:IBMJ}. The time scale separation in many realistic systems is not always satisfied, so that
we will relax the above local Born approximation
in Sec.~\ref{sec3b} and systematically include the contribution of bath relaxation.

In this subsection, we discuss higher-order corrections of system coherence,
i.e., multi-site coherence~\cite{cao2009:JPCA,JLWu2012:JCP}, under the local Born approximation. After averaging over bath in Eq.~(\ref{eq007}),
the time evolution equation of the system population is given by,
\be
\dot{P}(t) &=& - \Tr_b\left\{\left[\calW^{(2)}+\calW^{(3)}+\calW^{(4)}+\cdots\right]\star\rho^\eq_B\right\}\ast P,
\label{eq008}
\ee
where $\rho^\eq_B$ is the vector of the equilibrium bath density state at each local site, i.e.,
$\rho^\eq_B=\left(\rho^\eq_{b_1}, \rho^\eq_{b_2},\cdots\right)^T$.
Here the superscript $T$ denotes matrix transpose,
and the symbol $\star$ defines a matrix product, $(X\star Y)_{mn} = X_{mn}Y_n$, of a matrix $X$ and a vector $Y$.
To be consistent, each local bath is in equilibrium initially, i.e., $\rho_{\rmP; n}(0) = P_n(0)\rho^\eq_{b_n}$,
which belongs to the class-B preparation in Ref.~\cite{Weiss1999:book}.

For conciseness, we introduce two quantum bath operators,
$\langle X = \Tr_b\left\{X \right.$, and $ X\rangle = \left. X\star\rho^\eq_B\right\}$.
Consequently, we define the bath average, $\langle X\rangle = \Tr_b\left\{X \star\rho^\eq_B\right\}$, and
the projection onto the bath equilibrium density state, $\rangle \langle~ =\left. \star\rho^\eq_B\right\}\Tr_b\left\{ \right. $.
Equation~(\ref{eq008}) is simplified to a time-convolution form,
$\dot{P}(t) = -\langle \calW\rangle\ast P$.
Compared to the second-order truncation approaches such as Fermi's golden rule rate and NIBA,~
Eq.~(\ref{eq008}) systematically includes time-nonlocal corrections of multi-site quantum coherence,
$\{\calW^{(3)},~\calW^{(4)},~\cdots\}$, resulted from direct interconversion of coherence in the Liouville space.

\subsection{Bath Relaxation Effect}
\label{sec3b}

Normally, the bath requires a characteristic time scale to adjust to the system change
and relax back to equilibrium. The resulting memory kernel can be crucial
for long-lived quantum coherence in light-harvesting systems~\cite{Engel2007:Nature,Collini2009:Science}
%Harel2012:PNAS,Collini2010:Nature,Ishizaki2009:PNAS} %Collini2009:Science} and
and solvent-modified electron transfer reactions~\cite{Marcus1964:ARPC}.
Therefore, we need a systematic and reliable way to include the
contribution of bath relaxation beyond the Born approximation.
To compute higher-order corrections from both quantum coherence and bath relaxation,
we  extend an approach previously for the two-site system~\cite{Cao2000:JCP2} to the general $N$-site system.

Integrating the time differential equation in Eq.~(\ref{eq007}),
the total population vector in a non-equilibrium bath is written explicitly as
a time-convolution form, $\rho_\rmP(t) = \calU_\rmP(t) \rho_\rmP(0) -\calU_\rmP \ast \calW \ast \rho_\rmP$,
where $\calU_\rmP(t)$ is the time evolution matrix of population, $\calU_\rmP(t) = \exp(-i\calL_\rmP t)$.
For a product initial state, $\rho_{\rmP; n}(0)=P_n(0)\rho^\eq_{b_n}$,
the bath average leads to the system population in the form of
\be
P(t) &=& P(0) - \left[\ast \langle\calW^{(2)}\rangle \ast\right] P(0)
- \left[\ast \langle\calW^{(3)}\rangle\ast\right] P(0) \no \\
&&+\left[\ast\left(\langle\calW^{(2)}\ast\calU_\rmP\ast\calW^{(2)}\rangle-\langle\calW^{(4)}\rangle\right)\ast\right] P(0)
+ \cdots,
\label{eq009}
\ee
where the expansion order is the total number of the %three interaction terms,
Liouville superoperators, including $\calL^{(1)}_\rmC,~\calL_{\rmP\rmC}$, and $\calL_{\rmC\rmP}$.
Similar to that in Ref.~\cite{Cao2000:JCP2}, the notation of $[\ast X\ast]$ defines
a time convolution with the unit function in both the first ($\tau_1$)
and final ($\tau_k$) time steps,
i.e.,
\be
[\ast X\ast]=\int_0^t d\tau_1 d\tau_2\cdots d\tau_k  X(\tau_2, \cdots, \tau_{k-1})\de_{\tau_1+\tau_2+\cdots+\tau_k, t}.
\ee

On the other hand, we can formally assign a time-nonlocal kinetic equation,
\be
\dot{P}(t) = - \calK \ast P,
\label{eq010}
\ee
to describe the time evolution of system population $P$,
where $\calK$ is the time-nonlocal quantum rate kernel.
Similar to $\calW$,
the rate kernel $\calK$ can be expanded as $\calK = \calK^{(2)}+\calK^{(3)}+\calK^{(4)}+\cdots$,
in the order of the $\calL$ terms. %$\{\calL^{(1)}_\rmC,~ \calL_{\rmP\rmC/\rmC\rmP}\}$.
The integration of Eq.~(\ref{eq010}) then leads to
\be
P(t) =&& P(0) -  \left[\ast\calK^{(2)} \ast\right] P(0)-  \left[\ast\calK^{(3)} \ast\right] P(0)  \no \\
&&+  \left[\ast \left(\calK^{(2)} \ast \calK^{(2)}- \calK^{(4)}\right)\ast\right] P(0)+\cdots.
\label{eq011}
\ee
The term-by-term comparison between Eqs.~(\ref{eq009}) and~(\ref{eq011})
determines the explicit forms of the quantum rate kernels, e.g.,
\begin{subequations}
\be
\calK^{(2)} &= & \langle\calW^{(2)}\rangle,  \label{eq012a}   \\
\calK^{(3)}  &=& \langle\calW^{(3)}\rangle,  \label{eq012b} \\
\calK^{(4)}  &=& \langle\calW^{(4)}\rangle
-\left[\langle\calW^{(2)}(\tau_4)\calU_\rmP(\tau_3)\calW^{(2)}(\tau_2)\rangle-\calK^{(2)}(\tau_4)\calK^{(2)}(\tau_2)\right].
\label{eq012c}
\ee
\label{eq012tot}
\end{subequations}

\subsection{Higher-Order Quantum Rate Kernels and Kinetic Mapping}
\label{sec3c}

Extending the procedure in the previous subsection to higher orders,
we can straightforwardly derive the general form of the $k$-th rate kernel,
given by
\be
&&\calK^{(k>3)} (\tau_2, \cdots, \tau_k) = \langle \calW^{(k)}\rangle -\sum_{k_1, k_2\ge2}\de_{k_1+k_2, k}
\left[\left\langle \calW^{(k_1)}\calU_P\calW^{(k_2)}\right\rangle -\calK^{(k_1)}\calK^{(k_2)} \right] \no \\
& & + \sum_{k_1, k_2, k_3\ge 2}\de_{k_1+k_2+k_3, k}\left[\left\langle \calW^{(k_1)}\calU_P\calW^{(k_2)}\calU_P\calW^{(k_3)}\right\rangle
-\calK^{(k_1)}\calK^{(k_2)}\calK^{(k_3)} \right] +\cdots.
\label{eq013}
\ee
The right hand side of Eq.~(\ref{eq013}) is terminated
when each index $k_i$ of $\calK^{(k_1)}\calK^{(k_2)}\cdots\calK^{(k_i)}\cdots$ in the final
summation term is equal to either 2 or 3.
The summation terms subsequently changes between positive and negative signs.
The time variable sequence follows the same ordering, $\{\tau_2, \cdots, \tau_{k-1}, \tau_k\}$, in each
summation term. Comparing Eq.~(\ref{eq013}) with the expression in Sec.~\ref{sec3a}, we observe that in addition to
the correction  $\langle \calW^{(k)}\rangle$ from multi-site quantum coherence,
the bath relaxation (system-bath entanglement)
also influences quantum dynamics due to the  fluctuation
around the reference density state of the local Born approximation (the local equilibrium density state of bath).
For example, the second term of $\calK^{(4)}$
can be simplified to $\langle \calW^{(2)}\de\calU_P\calW^{(2)}\rangle$,
where $\de\calU_P=\calU_P-\rangle\langle$ represents the deviation from the local Born approximation.
Compared to the expression in Ref.~\cite{Mukamel1989:JCP}, Eq.~(\ref{eq013}) includes
the odd-$k$ terms from the imaginary accumulated phases.

The non-Markovian quantum kinetic equation in Eq.~(\ref{eq010}) together with
kinetic rate kernels in Eqs.~(\ref{eq012tot})-(\ref{eq013})
constructs a rigorous theoretical framework, i.e., the higher-order quantum kinetic expansion (QKE),
to compute the time evolution of the system population in a quantum network.
The quantum dynamics of the density matrix in the $N^2$-D Liouville space
is mapped onto kinetics in the $N$-D population space.
In the HSR model, such kinetic mapping was
developed based on the stationary approximation of coherence~\cite{cao2009:JPCA}.
In our current formalism, kinetic mapping is generalized for the arbitrary $N$-D quantum network
using the non-Makrovian rate kernel $\calK$.
The leading-order expansion, $\calK^{(2)}$, is the same as the rate kernel in the NIBA approach~\cite{Leggett1987:RMP}.
The time integration of $\calK^{(2)}$ recovers Fermi's
golden rule rate, which is often considered as the `classical' description of kinetics.
As corrections to $\calK^{(2)}$, higher-order rate kernels $\calK^{(k)}$ can
identify and quantify various nontrivial quantum coherent effects, which will be demonstrated by examples in Sec.~\ref{sec5}.

\subsection{Quantum Kinetic Rate Kernels Expressed in Hilbert Space}
\label{sec3d}

To compute quantum kinetic rate kernels, we express the superoperators
$\calL$ and $\calU(t)$ as functions of
 the Hamiltonian $H$ and the time evolution operator $U(t)$ in the Hilbert space.
Based on its definition, $\calL X \equiv [H, X]$,  the Liouville superoperator
$\calL$ is expanded to be, $\calL_{mn, kl}X_{kl} = H_{mk}X_{kl}\de_{n, l}-X_{kl}H_{ln}\de_{m, k}$,
where the positions of $H_{mk}$ and $H_{ln}$ are usually fixed
since these two Hamiltonian elements can be quantum operators of bath, the same for $X$.
In this paper, we ignore the fluctuation around off-diagonal Hamiltonian elements,
which leads to the following scalar forms,
\begin{subequations}
\be
\calL_{\rmP\rmC; m, kl}&=& J_{mk}\de_{m, l} - J_{lm}\de_{m, k},  \label{eq014a} \\
\calL_{\rmC\rmP; kl, m} &=& J_{km}\de_{l, m}-J_{ml}\de_{k, m},   \label{eq014b} \\
\calL^{(1)}_{\rmC; k_1l_1, k_2l_2}&=& J_{k_1 k_2}\de_{l_1, l_2}-J_{l_2 l_1}\de_{k_1, k_2}.
\label{eq014c}
\ee
\label{eq014tot}
\end{subequations}
The other two Liouville superoperators, $\calL_\rmP$ and $\calL^{(0)}_\rmC$, are diagonal in
the system basis. Each diagonal element of the two corresponding time evolution matrices behaves as
\begin{subequations}
\be
\calU_{\rmP; m}(t)X &=& U_m(t) X U^+_m(t), \label{eq015a} \\
\calU^{(0)}_{\rmC; mn}(t)X &=&  U_m(t) X U^+_n(t), \label{eq015b}
\ee
\label{eq015tot}
\end{subequations}
where $U_n(t)=\exp(-i H_n t)$ is the time evolution operator of the local site basis
$|n, b\rangle$ in the Hilbert space.

Next we substitute Eqs.~(\ref{eq014tot}) and (\ref{eq015tot}) into the expressions of quantum kinetic rate
kernels derived in Secs.~\ref{sec3b} and \ref{sec3c}.
For example, the second-order kinetic rate kernel becomes
\be
\calK^{(2)}_{mn(\neq m)}
= -2|J_{mn}|^2\Real~\Tr_b\left\{U^+_m(t)U_n(t)\rho^\eq_{b_n} \right\},
\label{eq016}
\ee
where `$\Real$' denotes the real part of a complex variable and the imaginary symbol `$\Imag$'
will also be used in this paper. The higher-order kinetic rate kernels can be similarly obtained.
So far our derivation is rigorous and general:
The surrounding bath can be an ensemble of harmonic or anharmonic oscillators with an arbitrary
spectral density. The bath can alternatively be defined by nuclear motion of
molecules and atoms, following
quantum or a classical dynamics.
%force field, e.g., the protein backbone and solvent
%molecules in natural light-harvesting systems.
The system-bath coupling $H_{SB}$ can
follow any functional forms in addition to the regular bilinear form.
For complex baths, numerical simulation will be required to
calculate the rate kernel of different orders.

\section{Higher-Order Quantum Kinetic Expansion for a Harmonic Bath}
\label{sec4}

In the remainder of this paper, we will focus on a blinear coupling between the system and
a harmonic (Boson) bath. %The other complex coupling forms could be further studied in the future.
In this section, we will derive analytical expressions of rate kernels
required in the higher-order QKE.
With the creation ($a^+_i$) and annihilation ($a_i$) operators for the $i$-th harmonic oscillator,
 %($\omega_i$) of the bath,
 the diagonal Hamiltonian element for each system site $|n\rangle$ is written as
\be
H_n =&& \e_n + \sum_i \omega_i a^+_i a_i + \sum_{i} \omega_i x_{ni} (a_i+a^+_i),
\label{eq017}
\ee
where the quantum zero-point energy $\omega_i/2$ is ignored and
the coefficient $x_{ni}$ is the system-bath coupling strength reduced by
the frequency $\omega_i$ of the $i$-th harmonic oscillator.
Quantum operators of different harmonic oscillators are assumed to
commute with each other, i.e., $[a^{(+)}_i, a^{(+)}_j]=0$ for $i\neq j$.
Equation~(\ref{eq017}) implicitly assumes a universal environment for all the system sites,
and an alternative approach is to apply an isolated environment for each site; these two methods
can lead to the same result.

\subsection{Canonical Transformation of the Displacement Operator}
\label{sec4a}

The trace of time-dependent operators over the quantum harmonic bath
can be solved by many theoretical techniques, e.g., the path-integral method~\cite{Feynman1963:AnnPhys,Weiss1999:book}.
Here we will apply a canonical transformation method
together with the cumulant expansion.

The displacement operator, $G_n = \exp\left[\sum_i x_{ni}(a^+_i-a_i)\right]$, is used to diagonalize
the bath-modulated diagonal Hamiltonian element, resulting in system-bath decomposition,
\be
G_n \hat{H}_n G^{-1}_n = \tilde{\vare}_n+H_B,
\label{eq018}
\ee
where a shift appears in the diagonal energy, $\tilde{\vare}_n = \vare_n  -\sum_i\omega_ix^2_{ni}$.
Although the same canonical operator is applied in the polaron method,
our general quantum kinetic equation formalism does not rely on the concept of polaron, as stated in Sec.~\ref{sec3}.
The diagonalization in Eq.~(\ref{eq018}) allows us to factorize
the local time evolution operator into a product form,
\be
U_n(t)  = \tilde{U}_{S; n}(t)[G^{-1}_n  U_b(t) G_n],  %e^{-i \tilde{\vare}_n t} [G^{-1}_n  U_b(t) G_n],
\label{eq019}
\ee
and express the local bath equilibrium state operator as
\be
\rho^\eq_{b_n} = G^{-1}_n \rho^\eq_b G_n,
\label{eq020}
\ee
where $\tilde{U}_{S;n}(t)=\exp(-i\tilde{\vare}_n t)$ is the time evolution function of the displaced system,
and the other two operators, $U_b(t)=\exp(-i H_B t)$ and $\rho^\eq_b \propto \exp(-\bt H_B)$,
only depend on bath. In the Heisenberg picture, the time-dependent displacement operator is then written as
\be
G_n(t) = U^+_b(t)G_nU_b(t) = \exp\left\{\sum_i x_{ni}[a^+_i(t)-a_i(t)]\right\},
\label{eq021}
\ee
where $a_i(t)=a_i e^{-i\omega_it}$ and $a^+_i(t) = a^+_i e^{i\omega_it}$
are time-dependent annihilation and creation operators, respectively.
Substituting the above results into the second-order quantum kinetic rate kernel $\calK^{(2)}$,
we arrive at
\be
\calK^{(2)}_{mn(\neq m)}(t)
&=& -2|J_{mn}|^2\Real~ e^{i\tilde{\vare}_{nm}t} \Tr\left\{ G_{nm}(t)G_{mn}\rho^\eq_b \right\} \no \\
&=& -2|J_{mn}|^2\Real~ e^{i\tilde{\vare}_{nm}t}\langle G_{nm}(t)G_{mn}\rangle_b,
\label{eq022}
\ee
with $G_{mn}=G_{m}G^{-1}_n = \exp\left[\sum_i x_{mn, i}(b^+_i-b_i)\right]$,  $x_{mn, i}=x_{mi}-x_{ni}$,
and $\tilde{\vare}_{nm}=\tilde{\vare}_{n}-\tilde{\vare}_{m}$.
The average, $\langle X\rangle_b=\Tr_b\{X \rho^\eq_b\}$, is taken over the
decoupled bath. By extending this method to higher-order expressions, we observe that all the quantum kinetic
rate kernels are fully determined by multi-time correlation functions of the canonical operator $G_{mn}$.

\subsection{Time Correlation Functions of Position Shift Operator}
\label{sec4b}

In this subsection, we derive the general form of multi-time correlation functions of $G_{mn}$ in the harmonic bath.

Applying the quantum thermal average of the harmonic bath, %to $G_{m_2 n_2}(t)G_{m_1n_1}$,
we obtain the analytical form of the two-time correlation function,
\be
\langle G_{m_2n_2}(t)G_{m_1n_1}(0)\rangle_b  = \exp\{-g_{m_2n_2, m_1n_1}(t)\},
\label{eq023}
\ee
where
\be
g_{m_2 n_2, m_1 n_1}(t) = \sum_{i} x_{m_1n_1, i}x_{m_2n_2, i}
\left[(1-\cos\omega_i t)\coth(\bt\omega_i/2)+i\sin\omega_i t\right].
\label{eq024}
\ee
In practice, we can assume a `spatial' correlation, $x_{mi}x_{ni}=c_{mn} x^2_i$, between each pair of
system sites, $|m\rangle$ and $|n\rangle$.
In the spin-boson model, a perfect negative correlation, $c_{mn}=2\de_{m,n}-1$, can be deduced
from the Pauli matrix $\sigma_z$~\cite{Leggett1987:RMP}. In energy transfer systems,
a zero spatial correlation, $c_{mn}=\de_{m,n}$, is often used~\cite{May2004}.
For a continuous bath, the spectral density is defined by $J(\omega) = \sum_i \omega^2 x^2_{i} \de(\omega-\omega_i)$,
and Eq.~(\ref{eq024}) is simplified to $g_{m_2 n_2, m_1, n_1}(t) = s_{m_1n_1,m_2n_2}g(t)$ with
\begin{subequations}
\begin{align}
s_{m_1 n_1, m_2 n_2}=& \left[c_{m_1m_2}+c_{n_1n_2}-c_{m_1n_2}-c_{m_2n_1}\right], \label{eq025a} \\
g(t) =&
\int_0^\infty d\omega [J(\omega)/\omega^2]\left[(1-\cos\omega t)\coth(\bt\omega/2)+i\sin\omega t\right].
\label{eq025b}
\end{align}
\label{eq025tot}
\end{subequations}
The above procedure can be straightforwardly extended to multi-time correlation functions following
the cumulant expansion of the Gaussian distributed noise.
In general, the $k$th-order time correlation function reads
\be
\left\langle G_{m_k n_k}(t_k)G_{m_{k-1} n_{k-1}}(t_{k-1})\cdots G_{m_1 n_1}(t_1)\right\rangle_b %\no \\&=&
=\exp\left\{-\left[\sum_{j=2}^k\sum_{j^\pr=1}^{j-1}g_{m_jn_j, m_{j^\pr}n_{j^\pr}}(t_j-t_{j^\pr})\right]\right\},
\label{eq026}
\ee
where the index set, $\{m_k, m_{k-1}, \cdots, m_1\}$, is the permutation of the original
set of indices, $\{n_k, n_{k-1}, \cdots, n_1\}$.
An additional constraint, $m_k=n_1$, is needed to close the index loop, as required by taking the trace.

\subsection{Three Leading-Order Quantum Rate Kernels in the Harmonic Bath}
\label{sec4c}

Through a tedious but straightforward derivation, we obtain the analytical forms of
quantum rate kernels in the harmonic limit. Here
we summarize and discuss the result of the three leading order rate kernels, which will be
applied to examples in Sec.~\ref{sec5}.

\begin{figure}[htp]
\includegraphics[width=0.8\columnwidth]{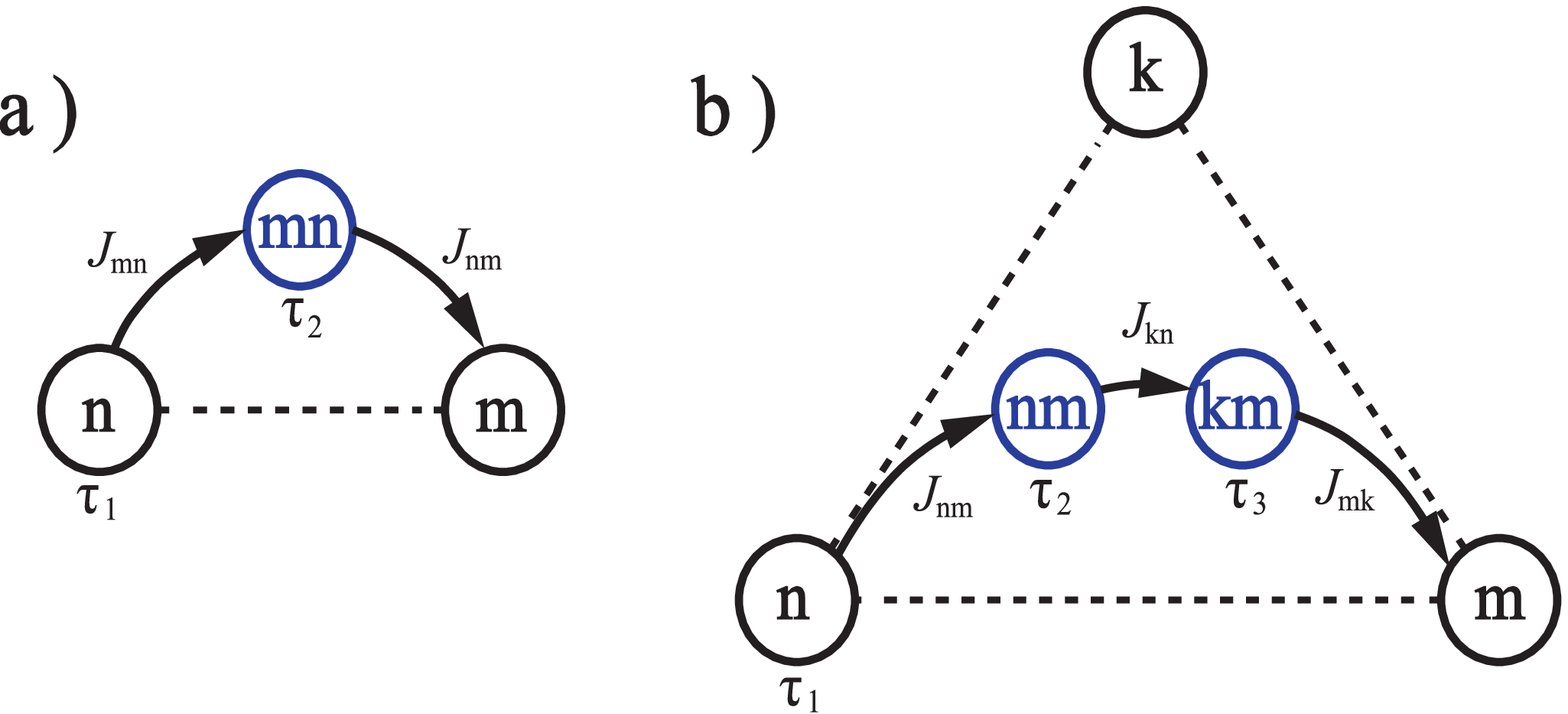}
  \caption{The diagrammatic representation of dynamic trajectories in
  the second- (a) and third-order (b) quantum rate kernels.
  Each circle with a single letter denotes a decoupled population state,
  e.g, $P_{n}(t)\rho^{\mathrm{eq}}_{b_n}$; each circle with two letters denotes a
  system-bath entangled coherence state, e.g., $\rho_{mn}(t)=\sum_{b, b^\pr}\rho_{mb, nb^\pr}(t)|b\rangle\langle b^\pr|$.
  The duration time, e.g., $\tau_{1, 2, \cdots}$,
  spanned at each density state (population or coherence) is given beneath its circle.
  Each dashed line between a pair of single-letter circles represents a nonzero
  interaction $J$ between these two sites in $H_{S}$. Each directed curve represents a transition
  from one density state to the subsequent one, where the inducing interaction, e.g., $J_{mn}$,
  is provided. Each integrated diagram composed of circles and
  connected curves describes one trajectory in the quantum rate kernels: a) a typical
  term in $\calK^{(2)}$ of Eq.~(\ref{eq027}) and b) a typical term in $\calK^{(3)}$ of Eq.~(\ref{eq028}). }
  \label{fig00}
\end{figure}

In the second-order quantum rate kernel, i.e., the NIBA rate kernel, each off-diagonal element is written as
\be
\calK^{(2)}_{mn(\neq m)}(\tau_2) &=&-2|J_{mn}|^2\Real \exp\left\{-[i\tilde{\vare}_{mn}\tau_2+s_{mn}g(\tau_2)]\right\},
\label{eq027}
\ee
with $s_{mn}=s_{mn, mn}=2(1-c_{mn})$.
The diagonal element $\calK^{(2)}_{nn}$ is
calculated by a summation, $\calK^{(2)}_{nn}=-\sum_{m(\neq n)}\calK^{(2)}_{mn}$.
Following the original equation of the total density matrix in Eq.~(\ref{eq005}),
we can interpret each term of the time convolution, $\calK\ast P$, as a dynamic
trajectory of density states in the Liouville space,
which determines the population evolution of
system at the next moment. The diagrammatic representation of
dynamic trajectories can clarify the effects of quantum coherence and
bath relaxation in each term of $\calK$.
Figure~\ref{fig00}a presents such a dynamic transition,
$P_{n}\rho^\eq_{b;n}\rightarrow \rho_{mn}\rightarrow P_{m}\rho^\eq_{b;m}$,
accounted in $\calK^{(2)}_{mn}$.
Here different circles of density states (population and coherence)
are connected by arrowed lines, representing the direction of the dynamic transition.
Each arrowed line is associated with a coupling $J$, which is the
interaction responsible for the transition
from one density state to the next one.
Our diagrammatic representation resembles the pathways in Ref~\cite{Mukamel1989:JCP},
but emphasizes the topology of the system Hamiltonian so that it is closer to kinetic mapping
representation in Ref.~\cite{cao2009:JPCA} and easier to extract different dynamic behaviors in
terms of expansion order. In addition, the factorized and unfactorized population states are
plotted together to highlight the reduced kinetics in the population subspace.

The higher-order rate kernels are corrections to $\calK^{(2)}$, related to
the multi-site quantum coherence and the bath relaxation.
In detail, the third-order rate kernel is given by
\be
\calK^{(3)}_{mn(\neq m)}(\tau_2, \tau_3) &=&
2~\Imag\left\{J_{mn}J_{n k}J_{km}e^{i(\tilde{\vare}_{nm}\tau_2+\tilde{\vare}_{km}\tau_3)-F^{-}_{3,kmn}}\right.\no \\
&&-\left.J_{n m}J_{mk}J_{kn}\left[
e^{i(\tilde{\vare}_{nk}\tau_2+\tilde{\vare}_{mk}\tau_3)-F^{-}_{3,mkn}}
+e^{i(\tilde{\vare}_{nk}\tau_2+\tilde{\vare}_{nm}\tau_3)-F^{+}_{3,mnk}}\right]\right\},
\label{eq028}
\ee
with $F^{\pm}_{3,abc} = s_{ca, cb}g(\tau_2)+s_{ab, ac}g(\pm\tau_3)+s_{ba, bc}g(\tau_2+\tau_3)$.
The summation over the extra system basis index $k$ is implied in
Eq.~(\ref{eq028}), and the same notation is applied to
the other higher-order rate kernels.
A typical dynamic transition,
$P_{n}\rho^\eq_{b;n}\rightarrow \rho_{nm}\rightarrow \rho_{km}\rightarrow P_{m}\rho^\eq_{b;m}$,
from the RHS of Eq.~(\ref{eq028}) is plotted in Fig.~\ref{fig00}b.
The nonzero prefactor, $J_{mn}J_{nk}J_{km}$,
requires a closed interaction loop in the system,
so that $\calK^{(3)}$ does not appear in
a 1-D chain model under the nearest neighbor interaction~\cite{Mukamel1989:JCP,Weiss1994:PRE}.
For complex interactions, quantum phase interference
can be significant in $\calK^{(3)}$, which will be demonstrated by an example
of the three-site system in Sec.~\ref{sec5d}.

\begin{figure}[htp]
\includegraphics[width=0.8\columnwidth]{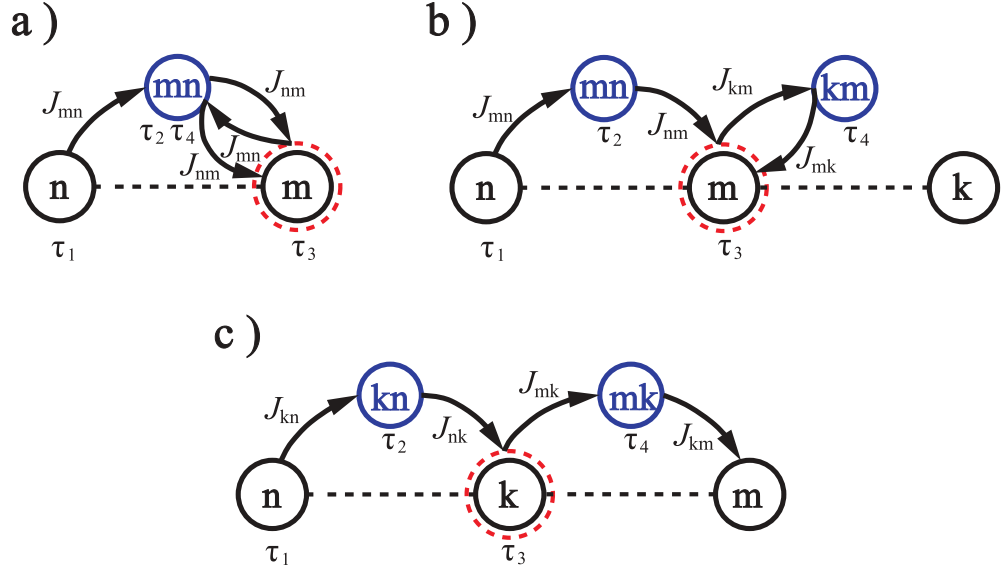}
  \caption{Three typical dynamic trajectories of density states in the fourth-order
  quantum rate kernel, $\calK^{(4)}_{\mathrm{bath}; mn}$.
  Each diagram represents a distinct behavior of the bath relaxation in Eq.~(\ref{eq029}) (see text).
  Here all the symbols, circles and lines have been explained explicitly in Fig.~\ref{fig00},
  except for the dashed circles that represent system-bath entangled population states,
  e.g., $\rho_{nn}(t)=\sum_{b, b^\pr}\rho_{nb, nb^\pr}(t)|b\rangle\langle b^\pr|$.}
  \label{fig01}
\end{figure}

The fourth-order quantum rate kernel can be divided into two terms depending on
the time evolution operator in the intermediate step:
$\calK^{(4)}_{\mathrm{bath}}$ due to the bath relaxation ($\de\calU_\rmP$)
and $\calK^{(4)}_{\mathrm{coherence}}$ due to the multi-site coherence ($\calU^{(0)}_\rmC$).
The first term  $\calK^{(4)}_{\mathrm{bath}}$  is explicitly written as
\be
\calK^{(4)}_{m n(\neq m); \mathrm{bath}}
= &&2\Real\left\{|J_{mn}|^2 |J_{mk}|^2\left[ e^{i(\tilde{\vare}_{nm} \tau_2 + \tilde{\vare}_{km} \tau_4 )}
e^{-[s_{mn}g(\tau_2)+s_{mk}g(-\tau_4)]} \left(e^{-s_{nm, km}F^{-}_{4A}}-1\right) \right. \right.\no \\
&&~~~~~~~\left. + e^{i(\tilde{\vare}_{nm} \tau_2 + \tilde{\vare}_{mk} \tau_4 )}
e^{-[s_{mn}g(\tau_2)+s_{mk}g(\tau_4)]} \left(e^{s_{nm, km}F^{-}_{4A}} -1\right) \right] \no \\
&&+|J_{mn}|^2|J_{nk}|^2 \left[  e^{i(\tilde{\vare}_{nk} \tau_2 + \tilde{\vare}_{nm} \tau_4 )}
e^{-[s_{nk}g(\tau_2)+s_{mn}g(\tau_4)]} \left(e^{-s_{mn, kn}F^{+}_{4A}}-1\right)\right. \no \\
&&~~~~~~~\left.+ e^{i(\tilde{\vare}_{nk} \tau_2 + \tilde{\vare}_{mn} \tau_4 )}
e^{-[s_{nk}g(\tau_2)+s_{mn}g(-\tau_4)]} \left(e^{s_{mn, kn}F^{+}_{4A}}-1\right)\right] \no \\
&&-|J_{mk}|^2 |J_{kn}|^2 \left[e^{i(\tilde{\vare}_{nk} \tau_2 + \tilde{\vare}_{mk} \tau_4 )}
e^{-[s_{nk}g(\tau_2)+s_{mk}g(-\tau_4)]} \left(e^{-s_{nk, mk}F^{-}_{4A}}-1\right) \right.\no \\
&&~~~~~~~\left.\left.+ e^{i(\tilde{\vare}_{nk} \tau_2 + \tilde{\vare}_{km} \tau_4 )}
e^{-[s_{nk}g(\tau_2)+s_{mk}g(\tau_4)]}  \left(e^{ s_{nk, mk}F^{-}_{4A}}-1\right)\right]\right\},
\label{eq029}
\ee
with
%\be
$F^{\pm}_{4A} =g(\pm\tau_3)-g(\tau_2+\tau_3)-g(\pm(\tau_3+\tau_4))+g(\tau_2+\tau_3+\tau_4)$.
%\label{eq048}
%\ee
As shown in Fig.~\ref{fig01}, the dynamic transitions in $\calK^{(4)}_{\mathrm{bath}}$
can be categorized into three types of diagrams:
a) The interaction prefactor is $|J_{mn}|^4$, and
the dynamic transition is within the sub-Liouville space of
the starting and ending system sites, $|n\rangle$ and $|m\rangle$.
A typical transitions is
$P_{n}\rho^\eq_{b;n}\rightarrow \rho_{mn}\rightarrow \rho_n \rightarrow \rho_{mn} \rightarrow P_m\rho^\eq_{b; m}$, % (Fig.~\ref{fig01}a),
where an intermediate population fluctuation occurs at site $n$
because the non-equilibrium bath is entangled with the system.
b) The interaction prefactor is $|J_{mn}|^2|J_{m(n)k}|^2$.
In the dynamic transition, a coherent state between $|m\rangle~(|n\rangle)$ and an additional site $|k\rangle~(\neq |m\rangle, |n\rangle)$ is involved
but the intermediate population fluctuation is still caused by the bath entangled with site $|m\rangle$ or $|n\rangle$.
A typical transition is
$P_{n}\rho^\eq_{b;n}\rightarrow \rho_{mn}\rightarrow \rho_m \rightarrow \rho_{mk} \rightarrow P_m\rho^\eq_{b; m}$. % (Fig.~\ref{fig01}b).
c) The interaction prefactor is  $|J_{mk}|^2|J_{nk}|^2$, so that the intermediate population fluctuation
is caused by the  bath entangled with the additional site $|k\rangle$.
A typical transition is $P_{n}\rho^\eq_{b;n}\rightarrow \rho_{nk}\rightarrow \rho_k \rightarrow \rho_{km} \rightarrow P_m\rho^\eq_{b; m}$. %(Fig.~\ref{fig01}c).
In the two-site system, only the first type of trajectories can appear~\cite{Cao2000:JCP2}.
In the $N$-site system, the bath relaxation can also induce a long-range transport
from the second and third types of trajectories in Figs.~\ref{fig01}b and c, in addition to the multi-site coherence.

\begin{figure}%[htp]
\includegraphics[width=0.8\columnwidth]{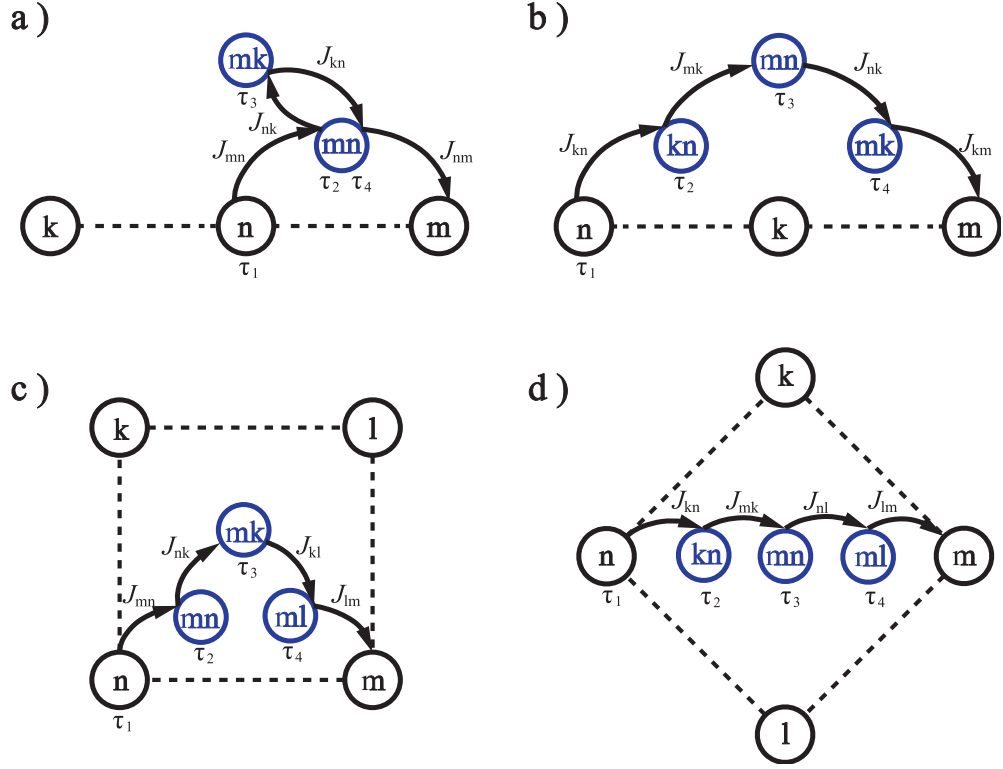}
  \caption{Four typical dynamic trajectories of density states in the fourth-order
  quantum rate kernel, $\calK^{(4)}_{\mathrm{coherence}; mn}$.
  Each diagram represents a distinct behavior of the multi-site quantum coherence in Eq.~(\ref{eq030}) (see text).
  Here all the symbols, circles and lines have been explained explicitly in Fig.~\ref{fig00}.}
  \label{fig02}
\end{figure}

The other fourth-order term, $\calK^{(4)}_{\mathrm{coherence}}$, due to the multi-site coherence is explicitly given by
\be
\calK^{(4)}_{\mathrm{coherence}; m n(\neq m)}&=& 2 \Real\left\{
J_{n k}J_{k l}J_{lm}J_{mn}e^{i (\tilde{\vare}_{nm} \tau_2+\tilde{\vare}_{km} \tau_3+\tilde{\vare}_{lm} \tau_4 )-F^{-}_{4B;kmnl} }
\right. \no \\
&&+ J_{m k}J_{k l}J_{ln}J_{nm}
\left[e^{i (\tilde{\vare}_{nl} \tau_2+\tilde{\vare}_{nk} \tau_3+\tilde{\vare}_{nm} \tau_4 ) -F^{+}_{4B;knlm}}  \right. \no \\
&& \left. +e^{i (\tilde{\vare}_{nl} \tau_2+\tilde{\vare}_{ml} \tau_3+\tilde{\vare}_{mk} \tau_4 ) -F^{-}_{4C;mlnk} }
 +e^{i (\tilde{\vare}_{nl} \tau_2+\tilde{\vare}_{nk} \tau_3+\tilde{\vare}_{mk} \tau_4 )-F^{+}_{4C;knlm} }\right] \no \\
&&-J_{m k}J_{kn}J_{n l}J_{lm}
\left[e^{i (\tilde{\vare}_{nk} \tau_2+\tilde{\vare}_{nm} \tau_3+\tilde{\vare}_{lm} \tau_4 )-F^+_{4C;mnkl} } \right.\no \\
&&\left.\left.+\left(e^{i (\tilde{\vare}_{nk} \tau_2+\tilde{\vare}_{lk} \tau_3+\tilde{\vare}_{lm} \tau_4 ) -F^{-}_{4C;lknm} }
+e^{i (\tilde{\vare}_{nk} \tau_2+\tilde{\vare}_{lk} \tau_3+\tilde{\vare}_{mk} \tau_4 )-F^{-}_{4B;lknm} }\right)\right]\right\},
\label{eq030}
\ee
with
\begin{subequations}
\be
F^{\pm}_{4B;abcd}&=&s_{ac, bc}g(\tau_2)+s_{ac, ad}g(\pm\tau_3)+s_{ad, bd}g(\pm\tau_4)\no \\
&&+s_{ad, cb}g(\tau_2+\tau_3)+s_{ac, db}g(\pm(\tau_3+\tau_4))+s_{bc, bd}g(\tau_2+\tau_3+\tau_4), \label{eq031a} \\
F^{\pm}_{4C;abcd}&=& s_{ac, bc}g(\tau_2)+s_{bd, ca}g(\pm\tau_3)+s_{bd, ad}g(\mp\tau_4) \no \\
&&+s_{bd, bc}g(\tau_2+\tau_3)+s_{da, ca}g(\pm(\tau_3+\tau_4)) +s_{da, bc}g(\tau_2+\tau_3+\tau_4). \label{eq031b}
\ee
\label{eq031tot}
\end{subequations}
Notice that the site indices, $m_1$ and $n_1$, for an arbitrary oscillation frequency, $\tilde{\vare}_{m_1 n_1}$,
 cannot be identical in Eq.~(\ref{eq030}).
Based on the number of additional system sites in $\calK^{(4)}_{\mathrm{coherence}}$, we identify two types of
multi-site coherence behaviors in the fourth order, and each type is further divided into two transition structures.
As shown in Fig.~\ref{fig02}a and b,  the first type of $\calK^{(4)}_{\mathrm{coherence}; mn}$ involves
one additional system site: a) The site $k$ interacts with either the starting ($|n\rangle$) or the ending ($|m\rangle$) site, and
the interaction prefactor is $|J_{mn}|^2|J_{n(m)k}|^2$. One example dynamic
transition is $P_{n}\rho^\eq_{b;n}\rightarrow \rho_{mn}\rightarrow \rho_{mk} \rightarrow \rho_{mn} \rightarrow P_m\rho^\eq_{b; m}$.
b) The site $|k\rangle$ interacts with both $|m\rangle$ and $|n\rangle$, and the interaction prefactor is $|J_{mk}|^2|J_{nk}|^2$.
One example transition is $P_{n}\rho^\eq_{b;n}\rightarrow \rho_{kn}\rightarrow \rho_{mn} \rightarrow \rho_{mk} \rightarrow P_m\rho^\eq_{b; m}$.
As shown in Fig.~\ref{fig02}c and d, the second type of  $\calK^{(4)}_{\mathrm{coherence}; mn}$
involves two additional system sites, $|k\rangle$ and $|l\rangle$,
which interact with both $|m\rangle$ and $|n\rangle$ and form a closed loop: c) The starting and ending sites, $|m\rangle$ and $|n\rangle$,
are interacted. One example transition is
$P_{n}\rho^\eq_{b;n}\rightarrow \rho_{mn}\rightarrow \rho_{mk} \rightarrow \rho_{ml} \rightarrow P_m\rho^\eq_{b; m}$
with the interaction prefactor $J_{mn}J_{nk}J_{kl}J_{lm}$.
d) The two sites, $|m\rangle$ and $|n\rangle$, are not interacted. One example transition is
$P_{n}\rho^\eq_{b;n}\rightarrow \rho_{kn}\rightarrow \rho_{mn} \rightarrow \rho_{ml} \rightarrow P_m\rho^\eq_{b; m}$
with the interaction prefactor $J_{mk}J_{kn}J_{lm}J_{nl}$.
The long-range quantum transport in the linear chain system is
explained by the first type of trajectories~\cite{Mukamel1989:JCP,Weiss1994:PRE},
whereas the four-site quantum interference is described by the second type of trajectories.

\section{Examples of the Higher-Order Quantum Kinetic Equation}
\label{sec5}

In this section, we apply the higher-order QKE
to four model systems, examining its validity and reliability.
To reduce the computation cost,
we introduce the Markovian approximation in the rate kernels.
The time evolution of the population of system is changed to
 $\dot{P}=-K P$, where $K= K^{(2)}+K^{(3)}+K^{(4)}+\cdots$ is the effective rate matrix
defined by the time integration of the rate kernel,
\be
K^{(k)} = \int_0^\infty \prod_{i=2}^k d\tau_i \calK^{(k)}(\tau_{2}, \tau_3, \cdots, \tau_k).
\label{eq032}
\ee
The second-order effective rate, $K^{(2)}$, recovers Fermi's golden rule rate.
The Markovian approximation ignores the short-time quantum oscillation
but can reliably describe the overall population dynamics.

\subsection{Kinetic Mapping in the Haken-Strobl-Reineker Model}
\label{sec5a}

The first example is the Haken-Strobl-Reineker (HSR) model~\cite{Haken1972:ZPhysik,Haken1973:ZPhysik,Silbey1976:ARPC,cao2009:JPCA},
where the bath is a Gaussian classical white noise.
%All the rate kernels
Without bath relaxation terms,
only multi-site quantum coherence contributes to higher-order corrections, i.e., $\calK^{(k)}=\langle\calW^{(k)}\rangle$.
A stationary coherence approximation was applied to derive kinetic mapping of the HSR model~\cite{cao2009:JPCA}.
Here we will demonstrate that higher-order QKE
leads to the exactly same result.

The spectral density of white noise, $J(\omega)=\Gamma\bt\omega/2\pi$,
together with the high-temperature approximation, $\coth(\bt\omega/2)\approx 2/\bt\omega$,
yields the time correlation function, $g(t)= \Gamma|t|/2$,
where the imaginary part is omitted under the consideration of the classical noise.
Without spatial correlation, $c_{mn}=\de_{m,n}$, the
second-order kinetic rate (i.e., Fermi's golden rule rate) is obtained as
\be
K^{(2)}_{mn}=K^{(2)}_{nm}=-|J_{mn}|^2 \frac{ 2\Ga_{mn}}{\Ga^2_{mn}+\tilde{\vare}_{mn}^2}.
\label{eq033}
\ee
which is the same as that derived in Ref.~\cite{cao2009:JPCA}.

All the higher-order corrections can be straightforwardly calculated by substituting the linear function
of $g(t)$ into expressions of $K^{(k)}$. To demonstrate the validity,
we examine the closed-looped three-site model.
Following Eqs.~(\ref{eq028}) and (\ref{eq032}),
the third-order correction from site 2 to site 1 is given by
\be
K^{(3)}_{12} &=& -2~\Imag\left\{\frac{J_{13}J_{32}J_{21}}{\tilde{\Ga}_{21}\tilde{\Ga}_{31}}
+\frac{J_{13}J_{32}J_{21}}{\tilde{\Ga}_{32}\tilde{\Ga}_{31}}
+\frac{J_{13}J_{32}J_{21}}{\tilde{\Ga}_{32}\tilde{\Ga}_{12}}\right\},
\label{eq034}
\ee
where $\tilde{\Ga}_{mn}=\Ga_{mn}+i\tilde{\vare}_{mn}$ is the complex dephasing rate.
Equation~(\ref{eq034}) is identical to the result derived in Ref.~\cite{cao2009:JPCA},
and the same conclusion is applied to all the other HSR systems.

\subsection{The Bath Relaxation Effect in a Two-Site System}
\label{sec5b}

\begin{figure}%[htp]
\includegraphics[width=0.8\columnwidth]{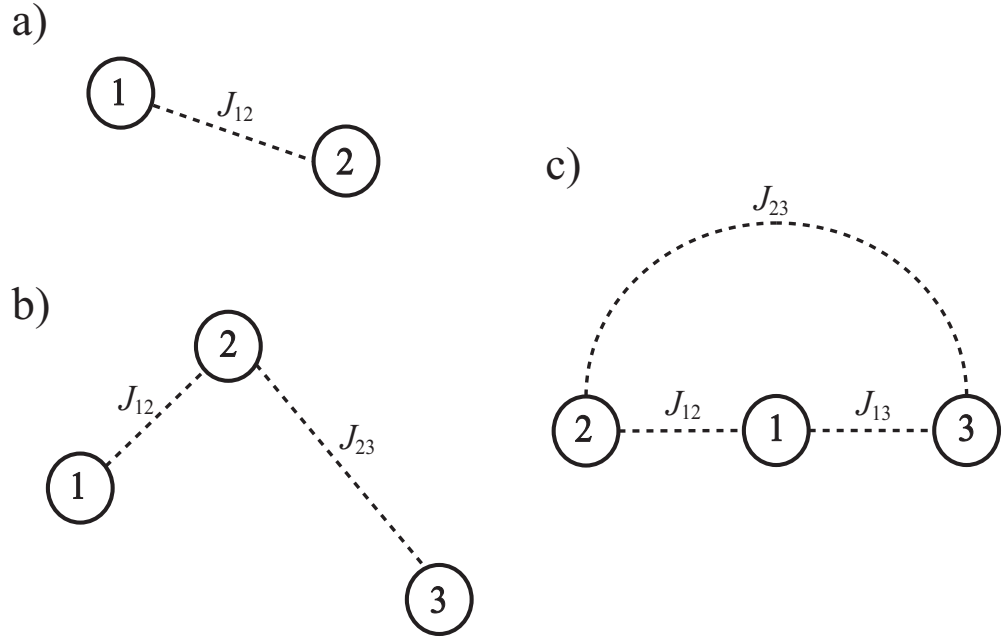}
  \caption{The schematic Hamiltonian diagrams of the three example quantum networks studied in Sec.~\ref{sec5}:
   a) the two-site system, b) the bridged three-site system, and c) the closed-looped three-site system.
   Here each circle with a number represents a `site' (basis) of the system. Each dashed line denotes
   a nonzero site-site coupling $J_{mn}$. The height of each circle denotes
   the relative energy $\varepsilon_{m}$ at each site. The closed-looped three-site system in c)
   actually forms a triangle network geometrically.
  }
  \label{fig03A}
\end{figure}

The second example is a two-site system in a quantum bath (see Fig.~\ref{fig03A}a). This model
is widely applied in the study of quantum transport
and quantum phase transition. % e.g., energy and charge transfers. %and the Kondo effect.
%a spin-boson model or a two-site exciton system with a single excitation.
%The change of the spatial correlation can be used to describe
%the variation from the spin-boson model~\cite{Leggett1981:PhysRevLett,Leggett1987:RMP}
%to the donor-acceptor pair in the energy transfer process~\cite{May2004}.
Without the coherence-coherence transition ($\calL^{(1)}_C=0$), %in the Liouville space,
all the terms with $\calW^{(k>2)}$ disappear and higher-order corrections only arise from the
bath relaxation effect, differing from the pure multi-site coherence effect in the HSR model.
In Ref.~\cite{Cao2000:JCP2}, The bath relaxation effect in the spin-boson model has been studied
following the short-time asymptotic expression of $g(t)$.
We will extend the calculation to the donor-acceptor pair with the zero spatial correlation, $c_{mn}=\de_{mn}$. To compare with the exact quantum dynamics,
we consider a quantum bath described by the Debye spectral density, which can be alternatively
solved by the hierarchy equation~\cite{Tanimura1989:JPSJ,Tanimura2005:JPSJ,Ishizaki2009:JCP2,Shao2004:CPL,Yan2005:JCP}.
%Ishizaki2009:JCP1,Ishizaki2009:PNAS,

The Debye spectral density is given by
\be
J(\omega) = \Theta(\omega) \left(\frac{2\lambda}{\pi} \right) \frac{\omega \omega_D}{\omega^2+\omega^2_D}
\label{eq035}
\ee
where $\Theta(\omega)$ is the Heaviside step function of $\omega$, $\lambda$ is the reorganization energy,
and $\omega_D$ is the Debye frequency. The inverse of $\omega_D$ represents the characteristic
time scale of bath relaxation, and the quantum coherence can be largely preserved as $\omega_D$ decreases.
To reduce the computation cost in the hierarchy equation, the high-temperature approximation is applied to
the time correlation function, resulting in
\be
g(t) \approx && \frac{2\lambda}{\bt\omega_D} \left[|t|-\frac{1-e^{-\omega_D|t|}}{\omega_D}\right]
+i~\mathrm{Sign}(t)\lambda\frac{1-e^{-\omega_D|t|}}{\omega_D},
\label{eq036}
\ee
where $\mathrm{Sign}(t)$ is the sign function of $t$.
In the short-time limit $(|t|\rightarrow 0)$, $g(t)$  asymptotically follows
$g(t) \sim \lambda t^2/\bt +i \lambda t$, which is applied in the study of electron
transfer~\cite{Cao2000:JCP2,Mukamel1989:JCP,Skinners1991:JCP,Reichman1996:JCP,Weiss1994:PRE}. %Reichman1997:PhysRevE,Mukamel1988:JPC,
In the long-time limit $(|t|\rightarrow \infty)$, the asymptotic time dependence becomes
$g(t) \sim 2\lambda|t|/\bt\omega_D+i~\mathrm{Sign}(t)\lambda/\omega_D$,
which resembles the result in the HSR model. To be consistent,
the high temperature approximation is also used in our higher-order QKE approach.

\begin{figure}[htp]
\includegraphics[width=0.85\columnwidth]{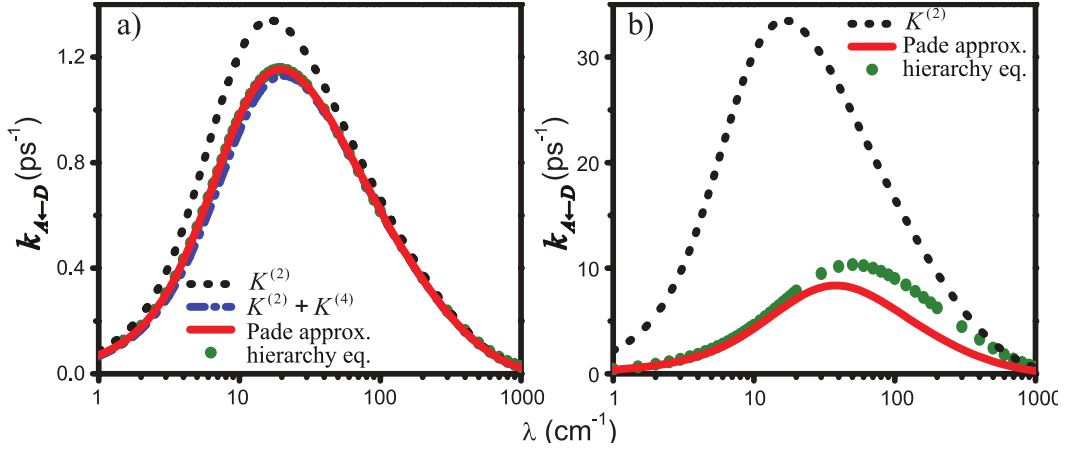}
  \caption{The effective forward rate ($k_{A\leftarrow D}$) from the donor to the acceptor in the two-site system (Fig.~\ref{fig03A}a)
  calculated using the higher-order QKE
  and the hierarchy equation.
  The detailed parameters are provided in Sec.~\ref{sec5b}. Among them, the site-site coupling is
  different in the two panels:  $J=20$ cm$^{-1}$ in a) and $J=100$ cm$^{-1}$ in b).
  The dashed line denotes the F\"{o}rster rate (i.e., the second-order rate of QKE).
  Both the dot-dashed and the solid lines include the fourth-order correction of bath relaxation,
 whereas the solid lines include the additional Pade approximation (see text).
  As a comparison, the data from the hierarchy equation are plotted as the solid dots. }
  \label{fig03}
\end{figure}

The parameters of our numerical calculation,
$\vare_{12}=100$ cm$^{-1}$,  $\omega^{-1}_D=100$ fs and $T=300$ K,
are taken from Ref.~\cite{Ishizaki2009:JCP2}. %Ishizaki2009:JCP1,
Two different site-site couplings, $J_{12}=20$ and 100 cm$^{-1}$, are chosen to test the
reliability of the higher-order QKE. The results of the effective forward rate ($k_{A\leftarrow D}$) from the donor to the acceptor from
the higher-order QKE and the hierarchy equation are plotted in Fig.~\ref{fig03}. For the small site-site coupling ($J=20$ cm$^{-1}$),
with the change of the reorganization energy $\lambda$,
a  difference up to $<40\%$ can be resolved between $k_{A\leftarrow D}$ calculated from
the hierarchy equation and from the second-order kinetic rate $k^{(2)}$ (i.e., the F\"{o}rster rate).
As a comparison, $k_{A\leftarrow D}\approx k^{(2)}+k^{(4)}$ after the leading order correction $k^{(4)}$
converges to the result of the hierarchy equation.  %in the whole regime of of $\lambda$.
With the Pade approximation~\cite{Mukamel1989:JCP,Cao2000:JCP2}, we apply the partial resummation technique, %Mukamel1988:JPC,
$k_{A\leftarrow D}\approx [k^{(2)}]^2/[k^{(2)}-k^{(4)}]$.
This improved prediction agrees perfectly with the result of hierarchy equation for an arbitrary $\lambda$.
To further demonstrate that the higher-order QKE is not limited in the regime of small site-site coupling,
we test a much larger value, $J_{12} = 100$ cm$^{-1}$, where the transportation does not follow the simple
hopping picture and the F\"{o}rster rate $k^{(2)}$ can be three times larger than the exact result.
Although $k^{(4)}$ causes an unphysical over-correction to %the F\"{o}rster rate
$k^{(2)}$, the prediction after the Pade approximation compares very well with the result of the hierarchy equation in the
whole regime of $\lambda$, especially in both coherent ($\lambda<20$ cm$^{-1}$) and incoherent
limits ($\lambda \sim 1000$ cm$^{-1}$).
By combining the higher-order QKE method with the
Pade approximation, the higher-order QKE  can thus improve the theoretical prediction of
quantum dissipative dynamics with a tolerable increased computation cost.

\subsection {Long-Range Energy Transfer in a  Three-Site Bridge System}
\label{sec5c}

In one model Hamiltonian of the seven-site FMO light-harvesting protein complex~\cite{JLWu2010:NJP,JLWu2012:JCP,Vulto1998:JPCB,Cho2005:JPCB},
the first energy transfer
pathway (sites $1\rightarrow 2\rightarrow 3$) carries a barrier crossing event at site 2,
and sites 1 and 3 are weakly coupled to each due to their long distance.
In classical kinetics, such a pathway is hindered by the barrier crossing from site 1 to site 2,
becoming less favorable compared to the alternative %longer
downhill pathway,
$6\rightarrow (5, 7)\rightarrow 4\rightarrow 3$.
However, our previous quantum-classical comparison~\cite{JLWu2012:JCP}
has showed that the first pathway can dominate even at the room temperature ($T=300$ K)
when the electronic excitation is initialized at site 1. The adjustment of
the energy transfer pathway is mainly caused by the direct energy transfer from site 1 to site 3
through multi-site quantum coherence.

%In the field of electron
%transfer, a similar behavior, the super exchange through bridge molecules, has been discussed
%for several decades~\cite{xxx}. In our GNIBA framework, these two phenomena can be unified.
%A super-exchange extended spin-boson model studied by Weiss {\it et. al.}~\cite{xxx} has been exactly
%recovered by us, which will not be shown in this paper due to the limiting space.

To demonstrate the long-range energy transfer phenomenon in a simple but transparent manner,
we select the sub-system of the first energy transfer path in our seven-site FMO
model.
We further set zero dipole-dipole coupling
between site 1 and site 3 %as an ideal condition
(see Fig.~\ref{fig03A}b) to neglect the irrelevant third-order correction $K^{(3)}$ but
focus on the leading-order corrections: the multi-site quantum coherence $K^{(4)}_{\mathrm{coherence}}$
and the bath relaxation $K^{(4)}_{\mathrm{bath}}$.
Following our previous papers~\cite{JLWu2010:NJP,JLWu2012:JCP}, the Hamiltonian of our three-site system is given by
\be
H_S = \left(\ba{ccc} 280 & -106 & 0 \\ -106 & 420 & 28 \\ 0 & 28 & 0 \ea\right)\mathrm{cm}^{-1}.
\label{eq037}
\ee
The Debye bath with the physiological condition is considered: $\lambda=35$ cm$^{-1}$,
$\omega^{-1}_D=50$ fs, and $T=300$ K, together with the zero spatial correlation, $c_{mn}=\de_{mn}$.

\begin{figure}%[htp]
\includegraphics[width=0.75\columnwidth]{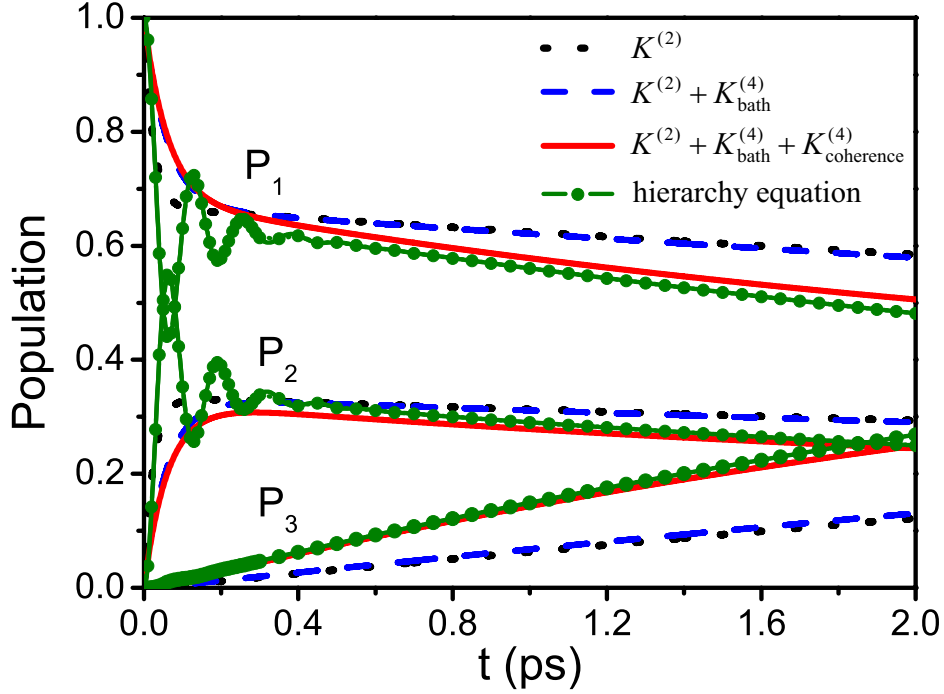}
  \caption{The population dynamics for the three-site bridge model (Fig.~\ref{fig03A}b)
  with the Hamiltonian in Eq.~(\ref{eq037})
  calculated by the higher-order QKE 
  with three different rate matrices  
  and by the hierarchy equation, respectively.  The bath parameters are provided in Sec.~\ref{sec5c}.
  From the top to the bottom, three distinct sets of curves represent the time evolution of $P_1(t)$, $P_2(t)$, and $P_3(t)$.
  Here the dotted lines are the results from $K^{(2)}$; the dashed lines are the results from $K^{(2)}+K^{(4)}_{\mathrm{bath}}$; the solid lines
  are results from $K^{(2)}+K^{(4)}_{\mathrm{bath}}+K^{(4)}_{\mathrm{coherence}}$. The Pade approximation is applied to
  all the fourth-order corrections. As a comparison, the results of the hierarchy equation
  are plotted   in the solid lines highlighted by the solid dots.}
  \label{fig04}
\end{figure}

The population dynamics predicted by the higher-order QKE
are plotted in Fig.~\ref{fig04}, together with the result of the hierarchy equation. %hierarchy equation.
We find that in the second order, the quantum kinetic equation using the F\"{o}rster rate
is unable to reliably predict both the short-time quantum oscillations and the long-time kinetics.
Following the Pade approximation,
the fourth-order corrections $K^{(4)}_{\mathrm{bath}}$ and $K^{(4)}_{\mathrm{coherence}}$
are included in the rate matrix.
With these leading-order corrections,
the prediction of the higher-order QKE is significantly improved, compared with that of the hierarchy equation. %hierarchy equation. %compared with the hierarchic result.
To avoid the overcorrection of these two terms, the %previous resummation from the
Pade approximation is used for each correction term in our calculation. Here we find that $K^{(4)}_{\mathrm{bath}}$ can
quantitatively describe the profile of slow-down ($<200$ fs)
in the time evolution of populations at sites 1 and 2,
although the exact quantum dynamics behaves as an under-damped oscillator.
Further improvement requires the non-Markovian  form of the time-nonlocal rate kernel $\mathcal K$.
Our comparison also determines that the bath relaxation does not affect the long-time
dynamics ($>200$ fs), possibly due to the fact that the bath relaxation time (50 fs)
is still much shorter than the overall energy transfer time ($\sim$ ps).
The more relevant multi-site coherence correction
$K^{(4)}_{\mathrm{coherence}}$ is shown to describe long-time population dynamics in a quantitatively reliable way.
We find that population accumulation at the trap site 3 is doubled compared to the prediction using the F\"{o}rster rate
in 2 ps. More importantly, we observe a direct evidence of the long-range energy transfer:
The majority of the fast increase in $P_3(t)$ after $t>200$ fs
arises from the decrease of $P_1(t)$ instead of $P_2(t)$,
which is consistent with the calculation of the flux network in the seven-site FMO model~\cite{JLWu2012:JCP}.
Overall, the nontrivial quantum coherent effect and the bath relaxation effect are identified and distinguished through
$K^{(4)}_{\mathrm{bath}}$ and $K^{(4)}_{\mathrm{coherence}}$.

\subsection{Quantum Phase Interference in a Closed Three-Site Loop Model}
\label{sec5d}

In additional to quantum tunneling, another unique and nontrivial quantum effect is
the interference of quantum phases. In this subsection, we use a closed three-site loop model as our
last example (see Fig.~\ref{fig03A}c)  to demonstrate the quantum phase interference
successful predicted by the higher-order QKE. %quantum kinetic equation.
The result of the classical white noise in Eq.~(\ref{eq034})
is extended to the quantum Debye noise,
and the fourth-order corrections are also included as a comparison.

For simplicity, we consider a degenerate three-site system with the following Hamiltonian,
\be
H_S = \left(\ba{ccc} 0 & J_{12} & J_{13} \\ J^\ast_{12} & 0 & J_{23} \\ J^\ast_{13} & J^\ast_{23} & 0 \ea\right).
\label{eq038}
\ee
All the site-site couplings are further assumed to be the same in the amplitude, $|J_{12}|=|J_{23}|=|J_{13}|=20$ cm$^{-1}$.
%but with different phases.
We assume all the other couplings are real positive and
check two phases for the coupling between sites 1 and 3: a) $J_{13}=20$ cm$^{-1}$, and b)
$J_{13}=20i$ cm$^{-1}$. The imaginary phase in the second case might be generated by
a coherent laser pulse.
As shown in the study of the HSR model~\cite{cao2009:JPCA},
the quantum phase interference can cause a significant difference in quantum dynamics,
 leading to the optimization of energy transfer with the variation of quantum phase.

\begin{figure}%[htp]
\includegraphics[width=0.85\columnwidth]{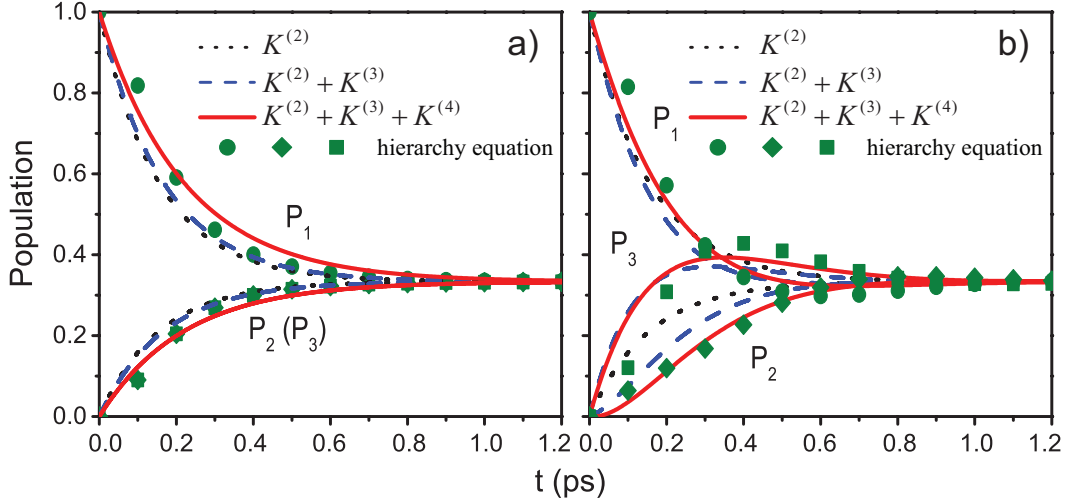}
  \caption{The population dynamics for the closed-looped three-site model
  (Fig.~\ref{fig03A}c) with the Hamiltonian in Eq.~(\ref{eq038})
  calculated by the higher-order QKE with rate matrices of three orders and by the hierarchy equation.
  The bath parameters are provided in Sec.~\ref{sec5d}. The left panel a) presents the results of $J_{13} = 20$ cm$^{-1}$;
  the right panel b) presents the results of $J_{13}=20i$ cm$^{-1}$. In each panel, the time evolution curves
  are  labeled by $P_{1,2,3}$ for the three sites. For the higher-order QKE,
  the dotted lines are the results from $K^{(2)}$; the dashed lines are the results from $K^{(2)}+K^{(3)}$; the solid lines
  are results from $K^{(2)}+K^{(3)}+K^{(4)}$. The Pade approximation is applied to
  all the higher-order corrections. As a comparison, the results of the hierarchy equation are
  highlighted by symbols (circles, diamonds, and rectangles for $P_1$, $P_2$, and $P_3$, respectively).}
  \label{fig05}
\end{figure}

Here we apply the Debye spectral density ($\lambda=50$ cm$^{-1}$ and $\omega^{-1}_D = 10$ fs)
together with a high temperature ($T=300$ K)
approximation to model the bath. The initial system is populated at site 1.
Under this particular reorganization ($\lambda=2.5|J|$),
it is expected that the short-time quantum oscillation is suppressed.
However, the nontrivial interference effect of quantum phase is still crucial
for the incoherent dynamics. The numerical results of
population dynamics using the higher-order QKE and the hierarchy equation
are plotted in Fig.~\ref{fig05} for both conditions of $J_{13}$.
In the second-order, Fermi's golden rule rate cannot distinguish the phase of $J_{13}$
and predicts the exactly same time evolution of population at sites 2 and 3.
As shown in Fig.~\ref{fig05}, the higher-order QKE %higher-order quantum kinetic equation
clarifies the effect of quantum phase interference.
Similarly, the Pade approximation is applied to every higher-order correction term.
For the real positive value of $J_{13}$ in Fig.~\ref{fig05}a, %($J_{13}=20$ cm$^{-1}$),
$K^{(3)}$ is nearly negligible and the dynamics of the sites 2 and 3 remains degenerate, $P_2(t)\equiv P_3(t)$,
after including $K^{(3)}$ and $K^{(4)}$ in the quantum kinetic equation.
For the the imaginary value of $J_{13}$ in Fig.~\ref{fig05}b, the third-order correction, $\calK^{(3)}$,
causes a significant change in dynamics: 1) the population transfer out of the initial site 1 is
accelerated; 2) the increase of $P_3(t)$  is much faster than $P_2(t)$ in the short time regime;
3) a short-time uni-directional energy transfer, $1\rightarrow 3\rightarrow 2$, is determined.
The above phenomena can be attributed to the constructive interference for site 3 and the destructive
interference for site 2. If site 3 is connected to a population sinker, the imaginary coupling of $J_{13}$
can yield a higher energy transfer efficiency, implying the optimization on quantum phase accumulation~\cite{cao2009:JPCA}.
The fourth-order correction in this second condition is much less relevant.
In addition, the predictions of the higher-order QKE %higher-order quantum kinetic equation
agrees well with the results of the hierarchy equation %hierarchic equation approach
for both conditions, which again confirms the reliability of our methodology.

\section{Summary and Discussions}
\label{sec6}

\subsection{Summary}
In this paper, we have applied a new formalism
to derive a higher-order quantum kinetic expansion (QKE) approach to study
quantum dissipative dynamics for a multi-`site' system.
After the integration of quantum coherence and the average over the local equilibrium bath,
we derived a closed non-Markovian quantum kinetic equation to describe the time evolution of system populations.
In this time-convolution equation, the kinetic rate kernel $\calK$ is rigorously
and systematically expanded in the order of the site-site coupling $J$, i.e.,
off-diagonal elements of the system Hamiltonian.
The second-order rate kernel $\calK^{(2)}$ recovers the result of the NIBA method,
and its time integration gives Fermi's golden rule rate. The higher-order corrections, $\calK^{(k>2)}$, include the contribution
from the multi-site quantum coherence (the direct coherence-coherence transition) and
the bath relaxation (the system-bath entangled population state).
For a harmonic bath,
the analytical expression of the kinetic kernel $\calK$ is obtained using
the displacement operator and the Gaussian cumulant expansion.
Our higher-order QKE approach is examined in four model systems to demonstrate its reliability.
In the Haken-Strobl-Reineker (HSR) model, the higher-order QKE
leads to the identical kinetic mapping previously derived by the stationary approximation of coherence~\cite{cao2009:JPCA}.
Under a quantum Debye noise, the prediction of the higher-order QKE together with the Pade approximation
agrees very well with the exact result of the hierarchy equation.

Compared to many other theoretical approaches of quantum dissipative dynamics, the
higher-order QKE can quantify higher-order corrections to
the second-order prediction, i.e., the Fermi's golden-rule expansion.
For example, the bath relaxation can
slow down the direct transfer from donor to acceptor, and the exact quantum transfer rate is
consistently smaller than the F\"{o}rster rate (see Fig.~\ref{fig03}).
For the three-site bridge model in
Sec.~\ref{sec5c}, the higher-order QKE predicts that
the bath relaxation slows down the short-time dynamics ($<200$ fs)
whereas the multi-site coherence speeds up the long-range transfer process afterwards (see Fig.~\ref{fig04}).
For the closed three-site loop model in Sec.~\ref{sec5d},
the quantum interference described by $K^{(3)}$ breaks the symmetry in the `classically'
incoherent dynamics (see Fig.~\ref{fig05}). All these examples have
confirmed that our higher-order QKE can clarify distinct nontrivial effects of
multi-site quantum coherence and bath relaxation,  which are crucial for
understanding nontrivial quantum effects.

The theoretical studies in this paper and our previous two papers~\cite{cao2009:JPCA,JLWu2012:JCP}
form a self-consistent methodology of quantum-classical comparison and kinetic mapping
for  quantum dissipative dynamics. Compared to kinetic mapping of
the HSR model in Ref.~\cite{cao2009:JPCA}, the expansion technique in this paper has been
extended from a classically white noise to an arbitrary quantum bath. As a result, the bath relaxation and the multi-site
coherence are unified in the same theoretical framework. Consistent with the quantum-classical comparison
in Ref.~\cite{JLWu2012:JCP}, the long-range energy transfer is now quantified in the detailed time evolution
and isolated from the short-time bath relaxation. In principle, the higher-order QKE
can be applied to an arbitrary quantum dissipative dynamic system for the investigation of
the nontrivial quantum effects.
%, which might be related to biological functions in natural
%light-harvesting systems.

\subsection {Discussions}

The calculation of the four examples in this paper shows the numerical accuracy from the higher-order QKE.
Using the hierarchy equation as the benchmark for the quantum Debye noise, the higher-order QKE can
provide reliable, sometime accurate, description for both the average
transfer rate and the detailed time evolution.
The quantum dynamics of the $N$-site system is described by the
time evolution of the reduced density matrix in the $N^2$-D Liouville space.
The dimensionality of the rate matrix for the hierarchy equation
grows roughly $\sim N^{h+2}$ with the hierarchy expansion order
$h$ under the high-temperature approximation for the Debye noise.
On the other hand, the rate kernel $\calK$ in the higher-order QKE is always restricted in
the $N$-D population subspace.
Although the computation time in our method increases with the expansion order similarly,
the Markovian approximation dramatically reduces the cost by changing
the time-nonlocal kernels into the average rate matrix. The Markovian higher-order QKE can
predict the overall features of time evolution.
In addition, the partial re-summation technique based on the Pade approximation
further accelerates the convergence of the rate matrix $K^{(k)}$.
For example, with a large site-site coupling
($J_{12}=|\varepsilon_{12}|$), the re-summation of the leading-order correction,
$K^{(4)}$, has already resulted in an almost quantitative prediction of the transfer rate.
Thus, the higher-order QKE promises the potential of predicting the quantum dissipative dynamics with
an acceptable computational cost.

In the higher-order QKE, the quantum dynamic system is defined as a general $N$-`site' network form.
The so-called `site' can be further generalized as any basis set of the system
Hilbert space. Thus, the higher-order QKE is not restricted in energy transfer and electron transfer,
but can extended to other quantum dynamic processes.
The surrounding bath is also defined generally in the higher-order QKE. Many
sophisticated deterministic or stochastic methods, such as the hierarchy equation, the SC-IVR, the QUAPI,
the polaron-based methods, the path integral Monte Carlo, etc.,
are based on a presumed harmonic bath, which is in general a good approximation under many conditions.
However, theoretical methods for the anharmonic environments is also highly required.
Since the expressions of the time-nonlocal rate kernels $\calK^{(k)}$
in Eqs.~(\ref{eq012tot}) and (\ref{eq013}) are independent of the bath model,
the formulation in the Hilbert space, e.g, Eq.~(\ref{eq022}),
can work as the starting point of studying quantum dissipative dynamics in a complicated
environment. Numerical implementation will be worth in the higher-order QKE along this direction.

Our current derivation needs an incoherent preparation for the initial product state,
which can be a  strong assumption considering the experimental designs,
e.g., a coherent laser pulse can generate different initial states.
With an additional expansion, the improvement of
including the initial quantum coherence can be derived in the future.
Together, the time evolution of quantum coherence needs to be
derived after the higher-order QKE.
A more important question is about the
expansion parameter of the higher-order QKE, i.e., the site-site coupling $J_{mn}$.
The different site-site couplings are also not necessarily on the same order of magnitude.
To solve this difficulty, we can introduce the sub-system concept and construct
the expansion based on the weak coupling between sub-systems.
For example, the multichromophore F\"{o}rster resonance energy transfer (MC-FRET)
rate theory~\cite{Sumi1999:JPCB,Jang2002:JCP,Cao2012_MCFRET2} can serve as the second-order prediction %Seogjoo2004:PRL
in the extended higher-order QKE of multiple sub-systems instead of multiple sites.
The higher-order corrections then help reveal nontrivial quantum effects beyond
the MC-FRET result. Overall, the higher-order QKE requires future improvements to solidify
its construction and extend its applications.

\begin{acknowledgements}
The work reported here is supported by the National Science Foundation (CHE-1112825) and DARPA grant (W911NF-07-D-004).
JC is partly supported by the MIT Center for EXcitonics.
JW acknowledges the support by the National Science Foundation of China (NSFC-21173185),
the Fundamental Research Funds for the Central Universities in China (2010QNA3041),
and Research Fund for the Doctoral Program of Higher Education of China (J20120102).
\end{acknowledgements}

%\bibliographystyle{apsrev}
%\bibliography{GNIBA_bib}

\newpage

\end{document}